\documentclass[10pt,journal,compsoc]{IEEEtran}

\AtBeginDocument{%
  \providecommand\BibTeX{{%
    \normalfont B\kern-0.5em{\scshape i\kern-0.25em b}\kern-0.8em\TeX}}}

\usepackage{CJKutf8}

\usepackage{color}

\usepackage{stmaryrd} 
\usepackage{amsmath} 
\usepackage{multirow} 
\usepackage{subfigure}
\usepackage{rotating} 
\usepackage{amssymb}
\usepackage{booktabs}
\usepackage{hyperref}
\usepackage{comment}
\usepackage{cite}
\usepackage{caption}
\usepackage{cleveref}

\newcommand{\x}[0]{{\bf{x}}}

\begin{document}

\title{Visual Acuity Consistent Foveated Rendering towards Retinal Resolution}

\author{Zhi Zhang, Meng Gai, Sheng Li*,~\IEEEmembership{Member,~IEEE} 
\IEEEcompsocitemizethanks{
\IEEEcompsocthanksitem Zhi Zhang is with the School of Software and Microelectronics, Peking University, China.\\ E-mail: \{zhangzhi543\}@stu.pku.edu.cn \\
Meng Gai, Sheng Li are with the School of Computer Science, Peking University, China.\\ E-mail: \{gaimeng\,$|$lisheng\}@pku.edu.cn \\
Sheng Li is also with the National Key Laboratory of Intelligent Parallel Technology. \\
\IEEEcompsocthanksitem Sheng Li is the corresponding author.
}
}

\IEEEtitleabstractindextext{
\begin{abstract}
Prior foveated rendering methods often suffer from a limitation where the shading load escalates with increasing display resolution, leading to decreased efficiency, particularly when dealing with retinal-level resolutions.
To tackle this challenge, we begin with the essence of the human visual system (HVS) perception and present visual acuity-consistent foveated rendering (VaFR), aiming to achieve exceptional rendering performance at retinal-level resolutions. Specifically, we propose a method with a novel log-polar mapping function derived from the human visual acuity model, which accommodates the natural bandwidth of the visual system. This mapping function and its associated shading rate guarantee a consistent output of rendering information, regardless of variations in the display resolution of the VR HMD. Consequently, our VaFR outperforms alternative methods, improving rendering speed while preserving perceptual visual quality, particularly when operating at retinal resolutions.
We validate our approach using both the rasterization and ray-casting rendering pipelines. We also validate our approach using different binocular rendering strategies for HMD devices. In diverse testing scenarios, our approach delivers better perceptual visual quality than prior foveated rendering while achieving an impressive speedup of 6.5$\times$-9.29$\times$ for deferred rendering of 3D scenarios and an even more powerful speedup of 10.4$\times$-16.4$\times$ for ray-casting at retinal resolution. Additionally, our approach significantly enhances the rendering performance of binocular 8K path tracing, achieving smooth frame rates.
\end{abstract}

\begin{IEEEkeywords}
foveated rendering, retinal resolution, visual acuity, shading rate, log-polar mapping
\end{IEEEkeywords}
}

\maketitle

\IEEEpeerreviewmaketitle

\IEEEraisesectionheading{\section{Introduction}}

\begin{figure*}
  \centering
    \subfigure[GT (23 fps)]{
    \begin{minipage}{5cm}
    \centering
    \includegraphics[scale=0.4]{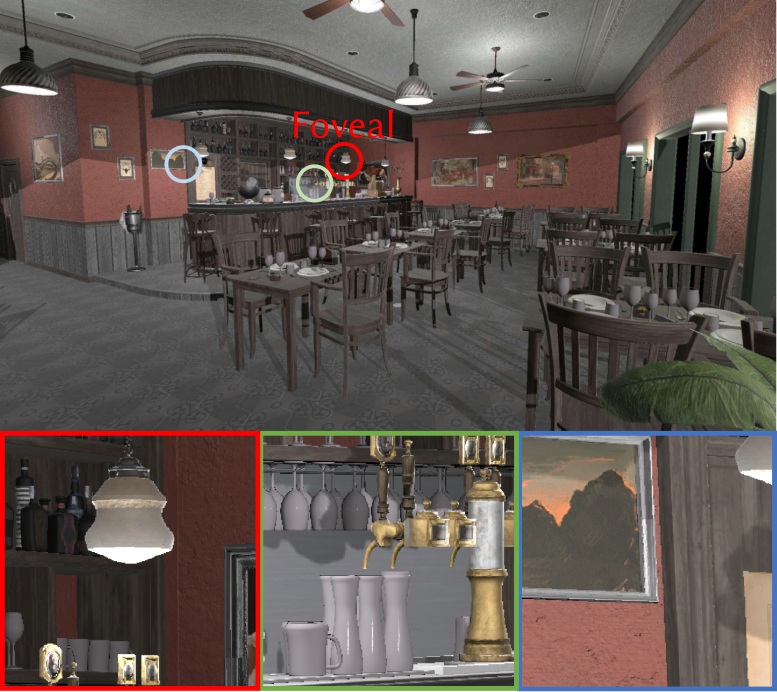}
    \end{minipage}
    }
    \subfigure[Our VaFR (210 fps)]{
    \begin{minipage}{5cm}
    \centering
    \includegraphics[scale=0.4]{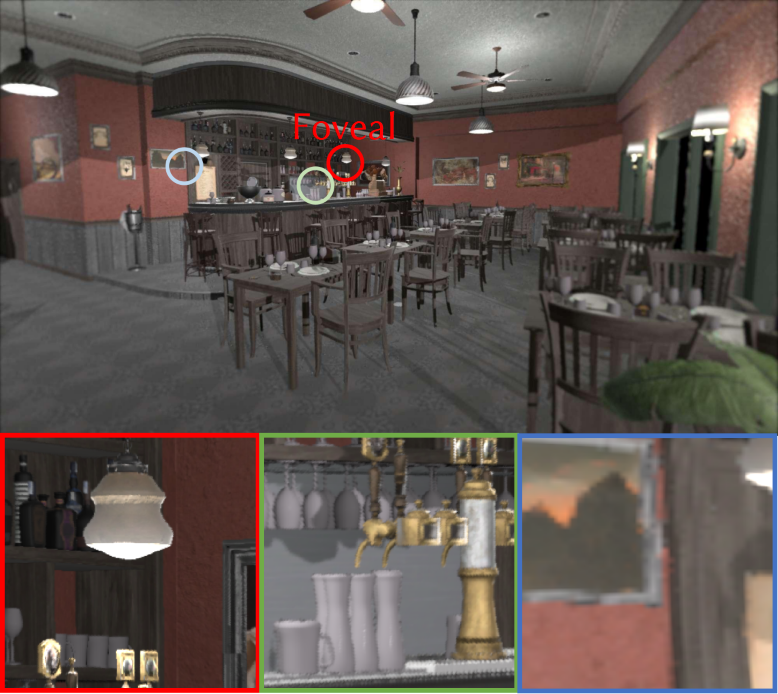}
    \end{minipage}
    }
    \subfigure[LaFR~\cite{Shi:2023} (49 fps)]{
    \begin{minipage}{5cm}
    \centering
    \includegraphics[scale=0.4]{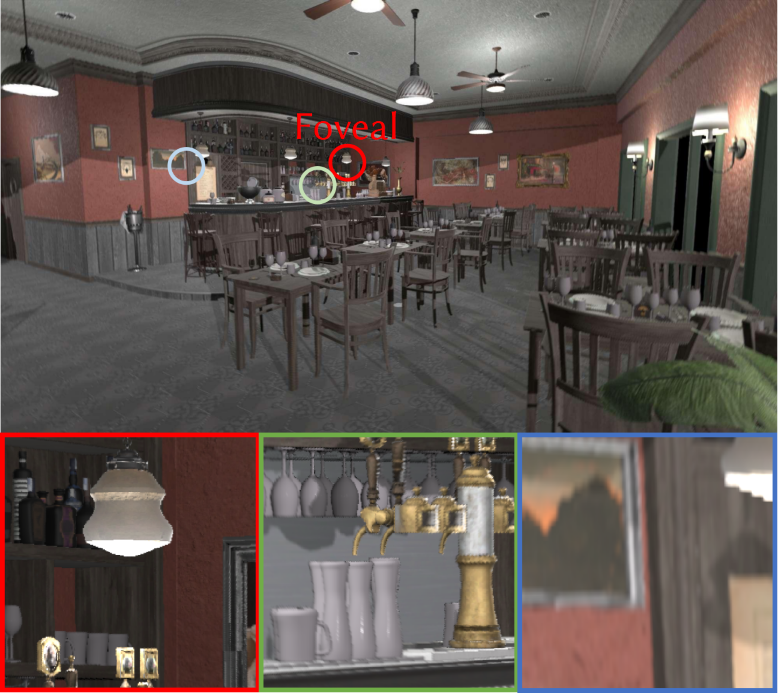}
    \end{minipage}
    }
  \caption{Results of our visual acuity consistent foveated rendering (VaFR) in the middle, ground truth (GT) on the left, and locomotion-aware foveated rendering (LaFR~\cite{Shi:2023}) on the right, rendered in ray casting mode at 8K ($7680\times4320$) resolution. Our approach, achieving an extraordinarily high frame rate of 210 fps for binocular rendering in VR settings, delivers comparable perceptual quality to that of LaFR (49 fps) and GT (23 fps).}
  \label{fig:teaser}
\end{figure*}

\IEEEPARstart{I}{n} virtual reality (VR) systems, rendering performance is critical for creating immersive experiences, with modern systems requiring high-quality rendering at frame rates above 90Hz \cite{Maimone:2017}. However, performance drops can disrupt immersion. Foveated rendering addresses these challenges by optimizing resource allocation based on human visual characteristics, thereby improving rendering efficiency. While the human visual field spans approximately $135^{\circ}$ vertically and $160^{\circ}$ horizontally, high visual acuity is concentrated in the fovea, a small central region covering just $1.5^{\circ}$ vertically and $2^{\circ}$ horizontally \cite{Guenter:2012}. By focusing computational power on this foveal region, foveated rendering enhances clarity where it is most needed and reduces resources allocated to peripheral areas, lowering overall computational demands. Additionally, real-time gaze tracking allows rendering engines to dynamically adjust detail levels, optimizing fidelity in critical areas.

Log-polar mapping, a transformation that converts Cartesian coordinate into a logarithmic-polar coordinate system, aligns the display space with human visual perception and can be used to approximate cortical excitation \cite{araujo1996introduction}.
The classic log-polar transformation for 2D image foveation on GPUs leverages the non-uniform distribution of visual acuity, emphasizing detail in the foveal region \cite{antonelli2015speeding}. As representatives, the log-polar mapping foveated rendering framework (LMFR) \cite{Meng:2018:Kernel} linearly reduces the resolution (inverse of the shading rate) outward from the fovea, employing a log-polar (LP) buffer for shading, while the locomotion-aware foveated rendering method (LaFR)~\cite{Shi:2023} dynamically adjusts the average shading rate based on user locomotion patterns.
However, these methods generally enhance performance only at lower resolutions (typically below 4K binocular display). As display resolution increases, the LP buffer size and shading load grow accordingly, leading to performance declines. Moreover, they rely on empirically determined model parameters attached to each HMD device’s resolution, often requiring manual tuning for optimal performance.

Human visual acuity near the fovea is extremely sharp, reaching an angular resolution of 1 arcminute (about 60 PPD), which in a $120^{\circ}$ field of view could equate to a resolution of 7200$\times$7200 pixels \cite{yang2023SIGTech}.
Achieving such high resolutions, even at retinal levels, poses significant challenges for prior methods. To address this, we introduce the visual acuity-consistent foveated rendering (VaFR) technique, designed for efficient rendering at the high resolutions of next-generation HMDs.

Our approach derives a novel log-polar mapping function based on human visual acuity, independent of the device display resolution. Using cortical magnification as a theoretical basis, we scale the log-polar transformation without relying on empirically determined model parameters. Additionally, we introduce variable scaling of the tangent plane in the log-polar coordinate system, ensuring that resolution on the log-polar tangent axis stays within human visual acuity limits.

Our VaFR decouples model parameters from device-specific display resolutions, delivering performance gains without compromising perceptual fidelity. By maintaining a constant-size LP buffer across resolutions, VaFR removes the need for resolution-specific parameter tuning required by previous methods, enabling consistent and efficient rendering aligned with human visual acuity. Additionally, our approach is adaptable to various VR rendering pipelines, achieving over 131 fps at $11520\times6480$ resolution per eye. 

In summary, our contributions include:
\begin{itemize}
\item We developed a visual acuity-based log-polar mapping function and derived optimal settings for highly efficient foveated rendering independent of device-specific display resolutions, eliminating the need for empirically determined model parameters.
\item We proposed a method for computing shading rates in log-polar algorithms, allowing independent adjustment of tangential and radial rates to align with visual acuity and assess perceptual compatibility.
\item Our method ensures efficient rendering across various VR displays, especially at near-retinal resolutions, paving the way for ultra-high-definition virtual reality experiences in the future.
\end{itemize}

\section{Related Work}

\subsection{Foveated Rendering}
\label{subsec:foveated}
Foveated rendering methods perform high-resolution sampling for foveal regions and significant user-noticeable areas and low-resolution sampling for peripheral regions.

\textbf{Rasterization Paradigm}: In recent years, due to growing 3D model complexity and larger virtual scenes, researchers are increasingly exploring foveated rendering and enhancing geometric mesh rasterization for multi-spatial resolution. Guenter et al. \cite{Guenter:2012} introduced foveated rasterization, finely controlling rendering quality across the foveal, transitional, and peripheral regions. Stengel et al. \cite{stengel2016adaptive} introduced a sampling method based on the visual acuity fall-off model, spatio-temporal and spatio-luminance CSFs, and subsequently integrated the sampling method into the deferred rendering pipeline. Turner et al. \cite{turner2018phase} aligned the rendered pixel grid with the content of the virtual scene during rasterization and upsampling to reduce motion artifacts. To enhance speed while maintaining perceptual quality, Patney et al. \cite{Patney:2016} presented a foveated shading system, drastically reducing shading computation while introducing an anti-aliasing algorithm for peripheral detail recovery. Friston et al. \cite{Friston:2019} proposed a unified foveated rasterization pipeline for HMDs, employing per-fragment ray casting for VR rendering efficiency. 
Foveated rendering has been diversified to address various aspects, including accelerating shadow rendering \cite{young2020optimized} and improving peripheral rendering quality with pixel reuse techniques through reprojection \cite{Franke:2020}, a variable-rate shading pipeline that partitions the image into tiles and dynamically modifies the shading precision to enhance the rasterization performance \cite{jindal2021perceptual}, post-processing methods \cite{Walton:2021}, and foveated rendering via rectangular mapping \cite{Ye:2022}.

\textbf{Ray Tracing Paradigm}: Ray tracing allows control over the number of rays per pixel, enhancing rendering quality and naturally supporting spatial multi-resolution rendering. Fujita et al. \cite{2014FoveatedRR} pioneered foveated ray tracing, using pre-computed sampling maps and K-nearest neighbor algorithms to reconstruct sparse samples for final image generation. Building on this, Weier et al.~\cite{Weier:2016} incorporated reprojection techniques, recycling ray samples from previous frames. Koskela et al.~\cite{Koskela:2017,Koskela:2018,Koskela:2020} introduced progressive Monte Carlo path tracing to accelerate rendering in regions of interest. Willberger et al. \cite{willberger2019deferred} introduced a hybrid path tracing method to accelerate global illumination in foveated rendering. Further advances include luminance-contrast-aware sampling for optimized ray tracing \cite{tursun:2019}, spatio-temporal reservoir resampling for efficient light sample selection \cite{bitterli2020spatiotemporal}, and adaptive selective sampling tailored for HMDs combined with foveated rendering \cite{Kim:2021}.

\textbf{Other Extensions}: Recent foveated rendering research explores various innovative approaches. Sun et al. \cite{sun2017perceptually} achieved perceptual quality by using fewer rays through 4D light field foveated rendering. Other advances include GAN-based methods to improve the quality of foveated images and videos in peripheral regions \cite{kaplanyan2019deepfovea}, as well as NeRF-based foveated rendering \cite{deng2022fov}.

Meng et al. \cite{Meng:2018:Kernel} introduced LMFR, which transforms screen-space coordinates into the log-polar space for efficient lighting and emphasizes the importance of the kernel function selection in improving peripheral perception and image quality. X. Shi et al. \cite{Shi:2023} extended LMFR with LaFR, dynamically adjusting parameters based on user locomotion patterns. Fan et al. \cite{fan2024scene} stretches the image after the variant log-polar mapping to fill the buffer and then stretch the mapped pixels based on the visual importance map of the scene.

Both LMFR and LaFR employ kernel functions to transform the Cartesian coordinates into a log-polar system, creating a lower-resolution log-polar (LP) buffer for shading. These methods rely on the heuristic selection of kernel functions and a parameter $\delta$, where $\delta^2$ represents the ratio of full-resolution screen pixels to LP buffer pixels, typically $\delta \approx 2$ \cite{Meng:2018:Kernel,Shi:2023} based on trial, error, and user studies. Shading occurs at a resolution approximately $\delta^2 \approx 4$ times lower than the native display resolution, making these approaches efficient for displays up to 4K per eye. However, as display resolution increases, so do the LP buffer size and shading load, particularly at retinal-level resolutions, resulting in significant performance drops and resource inefficiency. While these methods demonstrate empirical effectiveness, they lack a robust theoretical foundation.

We aim to develop more intuitive and mathematically sound kernel functions to enhance understanding of peripheral rendering. Our investigation examines how the mapping function affects shading point distribution, particularly in relation to human visual acuity. This highlights the importance of aligning the shading distribution with visual acuity for accurate and cohesive rendering. However, our research found that using kernel functions to redistribute shading points is impractical and inefficient. Consequently, we have moved away from traditional kernel functions.

\subsection{Visual Acuity}

The human visual system consists of key components, including the eye’s optical elements, retinal structures like photoreceptors, and neural processing in the visual cortex \cite{Strasburger:2011}. The differences between foveal and peripheral vision are mainly due to these elements. Peripheral vision has a lower image quality due to refractive lens effects \cite{Ferree1931REFRACTIONFT,Thibos:1987:Retinal,Thibos:1987:Vision}. Beyond $10^{\circ}$ eccentricity, visual signals experience defocus, astigmatism, and chromatic aberration. In the fovea, cones and ganglion cells connect one-to-one, preserving spatial detail, while in the periphery, they connect one-to-many, effectively filtering information.

Early studies of peripheral vision indicated that increasing the stimulus size as eccentricity results in a similar perceptual appearance. This phenomenon is linked to the relative neural volume across various regions, known as the cortical magnification factor, represented by $M$. In general, this factor scales as an inverse linear function, i.e. $ M^{-1} = M_0^{-1} \cdot(1+\alpha E) $, where $E$ denotes the eccentricity angle, $ M_0 $ represents foveal magnification \cite{Cowey:1974} and $\alpha$ is the slope of this linear function. In contrast, the non-linear magnification model was also investigated \cite{Rovamo:1979}. In the early stages, Levoy et al. \cite{levoy1990gaze} combined the ray casting method used for volume rendering with the visual acuity fall-off model. 
Some studies have integrated visual acuity fall-off models with vertex decimation for level-of-detail (LoD) \cite{luebke2001perceptually,ohshima1996gaze}. Guenter et al. \cite{Guenter:2012} simulated acuity reduction using layered rendering around the gaze point. Weier et al. \cite{Weier:2016} applied the model in ray tracing for HMDs, while Swafford et al. \cite{Swafford:2016} used it for multi-resolution ambient occlusion and tessellation. Tariq et al. \cite{Tariq:2022} enhanced adaptive-resolution foveated rendering by generating image details with spatial frequencies visible but not fully resolvable by the human eye.

Building upon the cortical magnification function, we aim to distribute shading points from the LP buffer in alignment with human visual acuity. Using the idea of log-polar mapping without employing the kernel function and the linear cortical magnification function, we will derive a mapping function that differs in structure from the traditional log-polar mapping approach.

\section{Visual Acuity consistent Foveated Rendering}

First, we analyze the fundamental deficiencies of the prior methods as preliminary (\Cref{subsec:deficiency}). As a solution, we leverage human visual acuity models to derive a novel log-polar mapping function (\Cref{section:Formulation}), which maintains a constant-sized LP buffer irrespective of resolution variations, providing a constant stream of visual information consistent with the finite bandwidth of the human brain.
Following the segmented nature of the mapping function, a shading rate adaption is proposed to automatically adjust the shading rate for different devices and for further custom tuning (\Cref{subsection:segmented mapping}).
Lastly, we present the pipeline of our algorithm in \Cref{Framework}. 

\subsection{Preliminary}
\label{subsec:deficiency}
LMFR \cite{Meng:2018:Kernel} and LaFR \cite{Shi:2023} aimed to reduce shading load by lowering the LP buffer resolution. They used an empirical mapping function, fine-tuning the $K$ function and parameters to adjust the pixel distribution and optimize VR visuals. However, as later analysis shows, their approach essentially tries to fit the shading rate to the human visual acuity model, a case of putting the cart before the horse.

Specifically, the initial step of LMFR and LaFR involves converting the screen-space frame buffer pixel coordinate $(x,y)$ into $(u,v)$ in the LP buffer in the log-polar coordinate system as:
\begin{equation}
\begin{cases}
	u = K(\frac{\ln \sqrt{(x-x_0)^2+(y-y_0)^2}}{L}) \cdot w\  , \\
	v = (\frac{1}{2\pi}\tan^{-1}\frac{y-y_0}{x-x_0} + 1[y-y_0<0]) \cdot h\  .
\end{cases}
\label{K:theta}
\end{equation}
Here, $(x_0,y_0)$ represents the position of the gaze point in the screen space, $L$ denotes the longest log distance from the gaze point to the corner of the screen, $(w,h)$ denotes the size of the LP buffer, $K(\cdot)$ denotes the kernel function, and $1[\cdot]$ serves as the indicator function. $u$ denotes the horizontal ordinate of the LP buffer, while $v$ represents the vertical ordinate. However, importantly, the kernel function $K(z) = z^\alpha$ in LMFR \cite{Meng:2018:Kernel} and $K(z) = \frac{z^\alpha}{1-\alpha}$ in LaFR \cite{Shi:2023} rely on empirical choices, limiting the flexibility of the mapping function and the broader applicability of their methods. Here, $z$ is short for $\frac{\ln \sqrt{(x-x_0)^2+(y-y_0)^2}}{L}$ and $\alpha$ is an empirically determined model parameter. \Cref{Inconsistency under Different Resolutions} and \Cref{Inconsistency under Gaze Direction Changing} will further elaborate on this issue, highlighting that dynamically aligning shading rates with human visual acuity under changing VR conditions presents even greater challenges.

\subsection{Our Formulation}
\label{section:Formulation}

We establish a theoretical foundation for precise shading rate alignment in foveated rendering by deriving a mapping function directly from visual acuity. This ensures a strict correlation between shading rate distribution and visual acuity. Specifically, our mapping functions (\autoref{func:u(e)} and \autoref{func:v(e,theta)}) rigorously align both radial and tangential shading rates with the acuity model, optimizing rendering workload distribution. While prior methods also seek to optimize shading rate distribution to improve efficiency, they often over- or under-allocate shading rates in certain regions, leading to unnecessary resource consumption. By contrast, our approach effectively eliminates these inefficiencies, ensuring the optimal utilization of rendering resources in terms of shading rate.

When expressed in terms of the minimum angle of resolution (MAR), the reciprocal of visual acuity, a linear model aligns well with anatomical data, such as receptor density and performance outcomes in numerous low-level vision tasks \cite{Weymouth:1958, Strasburger:2011}. Acuity models serve as the foundation for the widely recognized theory of cortical magnification or M-scaling, often using a linear model based on the reciprocal of the visual cortex tissue volume allocated to each eccentricity slice.

\subsubsection{Mapping Function}
\label{sec:mapping function}
\begin{figure}[tb]
	\centering 
	\includegraphics[width=\columnwidth]{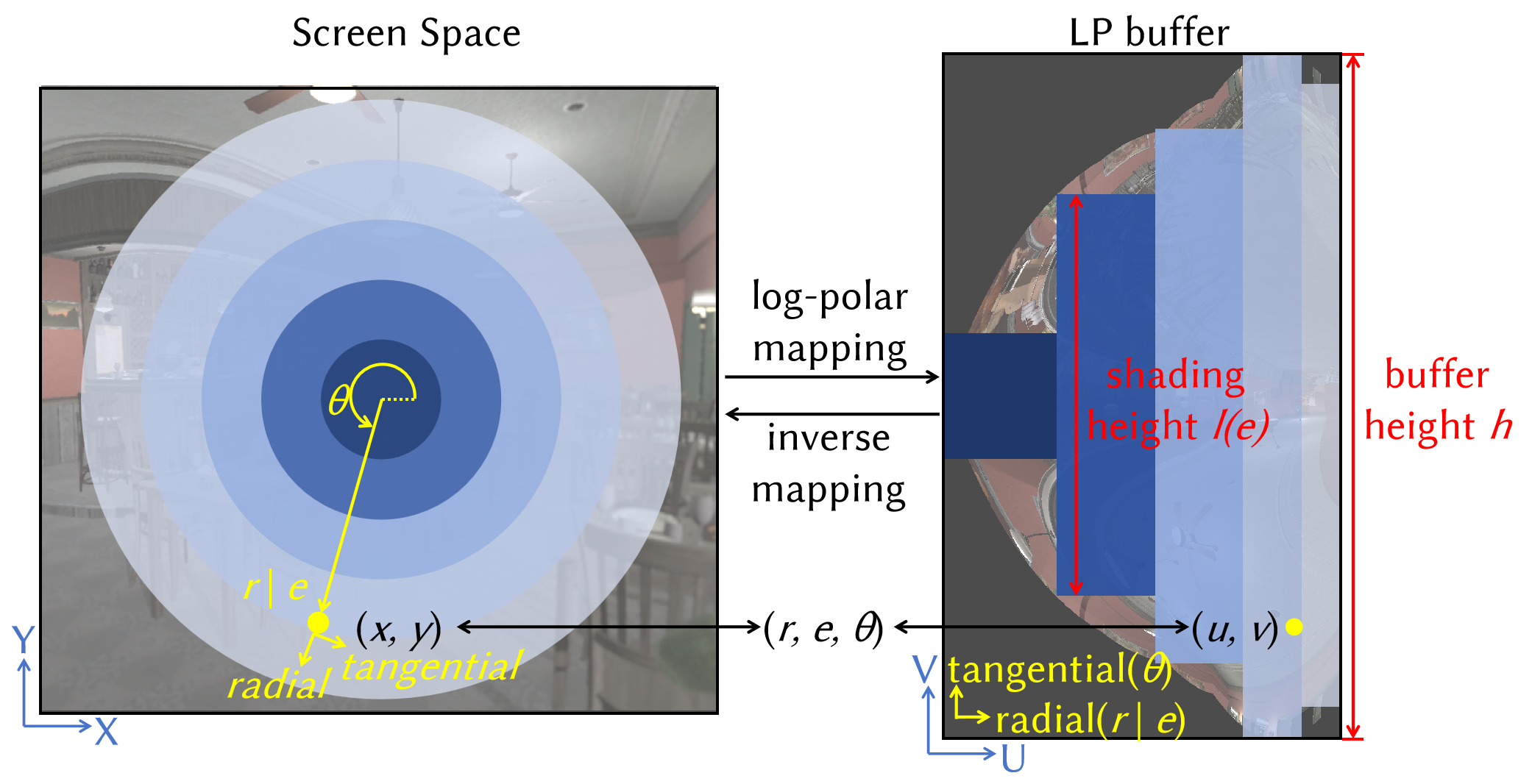}
	\caption{Schematic diagram of log-polar mapping and its inverse between the screen space and the log-polar space associated with the LP buffer, along with key variables in \Cref{sec:mapping function}. Additionally, we highlight the variable shading height $l(e)$, which differs from the buffer height $h$}.
	\label{fig:ShadingHeight}
\end{figure}

We illustrate the log-polar mapping and its inverse mapping between the screen space and the LP space in \autoref{fig:ShadingHeight}, along with the key variables. To derive the mapping function, we begin with the linear model:
\begin{equation}
	\omega(e) = m e+ \omega_0\  ,
\label{func:omega}
\end{equation}
where $\omega$ denotes MAR measured in degrees per cycle, $e$ refers to the eccentricity angle in degree, $\omega_0$ stands for the smallest resolvable angle that corresponds to the reciprocal of visual acuity at the foveal region (e = 0), and $m$ signifies the slope of the MAR. $f$ is acuity defined as the reciprocal of $\omega$ as $f = \frac{1}{\omega}$, expressed in cycles per degree (cpd) with each cycle corresponding to two pixels. 

To determine a constant ratio $c_r$ for a known camera and viewport in rendering, divide the camera's film plane height (in millimeters) by its focal length (also in millimeters). Then, divide this result by the display height in pixels of the native display resolution. The resulting ratio, $c_r$ represents the reciprocal of the pixel length corresponding to the focal length and simplifies the conversion equation between $e$ and $r$. It can be used in \autoref{func:r}:
\begin{equation}
	r = \frac{\tan e}{c_r} = \sqrt{(x-x_0)^2+(y-y_0)^2}\  ,
\label{func:r}
\end{equation}
where $e$ denotes the eccentricity angle in degrees, while $r$ refers to the pixel distance from any given point $(x,y)$ to the gaze point $(x_0,y_0)$. Therefore, we can infer: $e = \tan^{-1}(c_r \cdot r)$ from pixel distance $r$. Importantly, note that all trigonometric functions in the context are measured in degrees.

We have $f(e)$ from \autoref{func:omega} as: $f(e) = \frac{1}{\omega(e)} = \frac{1}{me+ \omega_0}$. Subsequently, each pixel in the log-polar space corresponds to half a cycle. Consequently, to ensure proper perception, each unit eccentricity angle should be shaded by $2f(e)$ pixels. Furthermore, $\frac{\mathrm{d}u}{\mathrm{d}e}$ indicates how many pixels in the LP buffer correspond to one unit of eccentricity angle, expressed as: 
\begin{equation}
	\frac{\mathrm{d}u}{\mathrm{d}e} = 2f(e) =  \frac{2}{m e+ \omega_0} \  .
\label{func:du/de}
\end{equation}
The significance of this equation is that each particular pixel in the LP buffer is mapped to a specific eccentricity angle, regardless of changes in gaze orientation or display resolution. This characteristic is a fundamental distinction between our algorithm and others. 
Accordingly, the shading rate can be defined as the actual number of shaded cycles in the log-polar space per unit of eccentricity angle. Specifically, the radial shading rate is the number of shaded cycles per unit of radial eccentricity angle, and the tangential shading rate is the number of shaded cycles per unit of tangential eccentricity angle.

Upon solving this equation, the horizontal coordinate $u$ in the LP buffer can be computed as follows:
\begin{equation}
	u(e) = 2\frac{\ln (me+\omega_0)}{m} + c_0\  .
\label{func:u(e)}
\end{equation}

To ensure the accuracy of the tangential shading rate, our algorithm initially set the tangential equal to the radial shading rates. Therefore, we can infer:
\begin{equation}
	r \frac{\mathrm{d}\theta}{\mathrm{d}v} = \frac{\mathrm{d}r}{\mathrm{d}u}\  .
\label{func:tangential}
\end{equation}
Two sides of this equation denote the ratio of the tangential pixel length in the screen space to the vertical pixel length in the LP buffer, and the ratio of the radial pixel length in the screen space to the horizontal pixel length in the LP buffer, respectively. For detailed information on the adjustment of the tangential shading rates by timing a ratio factor $\Delta(e)$, which varies with the eccentricity angle $e$ (see \Cref{Anisotropic}). 

By differentiating \autoref{func:r}, we can get:
\begin{equation}
	\frac{\mathrm{d}e}{\mathrm{d}r} = c_r\ \cos^2 e \cdot \frac{180}{\pi}\  .
\label{func:de/dr}
\end{equation}
Upon solving \autoref{func:tangential} and \autoref{func:de/dr}, the vertical coordinate $v$ in the LP buffer can be computed as:
\begin{equation}
	v(e,\theta) = \theta \cdot \frac{\sin{2e}}{me+\omega_0}\  .
\label{func:v(e,theta)}
\end{equation}
Next, we introduce the term shading height $l$ in the LP buffer, which refers to the varying number of shaded pixels along the vertical direction, as shown in \autoref{fig:ShadingHeight}. In previous studies, every pixel in each column of the LP buffer was shaded, meaning the shading height was always equal to the buffer’s total height. This rigid approach resulted in inaccuracies in the tangential shading rate and unnecessary rendering resource consumption.
In contrast, our method allows the shading height to adapt to the eccentricity angle $e$, as illustrated in \autoref{fig:ShadingHeight}, the total height of shaded pixels in each vertical column adjusts accordingly, reflecting how shading height evolves with changes in the eccentricity angle. 

Thus, the shading height at any given eccentricity angle $e$ is:
\begin{equation}
	l(e) = v(e,360^{\circ}) = 360 \cdot \frac{\sin{2e}}{me+\omega_0}\  .
\label{func:l(e)}
\end{equation}
This adaptability enables fine-grained control over the tangential shading rate, setting our algorithm apart from existing approaches. Details of the tangential shading rate in prior methods are presented in \autoref{fig:OurRadialTangential}, with their shortcomings discussed in \Cref{Anisotropic}. Unless specifically stated otherwise, all shading rates discussed in the following refer to radial shading rates.

\subsubsection{Constant-sized LP Buffer}
Assuming the maximum perceivable eccentricity as $e_{max} = 60^{\circ}$, a fixed LP buffer size of $(w,h) = (u_{max},l_{max}) = (u(60^{\circ}),v(e,360^{\circ})_{max}) = (932, 1800)$ is determined, where the shading pass is executed. Note that, we calculate the derivative of $l(e)$ equals zero, obtaining $e' \approx 45^{\circ}$ which maximizes $l(e)$, and then obtain $l_{max} \approx 1800$. From \autoref{func:u(e)} and \autoref{func:l(e)}, variables $u$ and $v$ cause the shaded image to shape as a semi-ellipse, as illustrated in the LP buffer in \autoref{fig:VaFR_framework}.

The variables $u$ and $l$ in our algorithm are uniquely linked to the variable $e$, while all other variables are set before rendering. This ensures a consistently sized LP buffer that remains unaffected by resolution or parameter changes. In contrast, both the LMFR and LaFR algorithms rely on empirical fine-tuning through user experiments at specific resolutions to determine the LP buffer size. Since their LP buffer size depends on the display resolution and parameter $\delta$, these algorithms can experience erratic shading rate fluctuations under changing conditions. Detailed issues will be discussed in \Cref{semantic}. 

A key feature of our algorithm is that its shading rate is solely determined by human visual acuity, unlike other methods that approximate this rate through coarse, resolution-specific tuning to fit the visual acuity model.
This unique characteristic, absent in prior algorithms, ensures a stable shading load across various display resolutions. Consequently, even at extremely high resolutions, such as $11520\times6480$ pixels per eye, our method can maintain robust performance while preserving retinal details in the foveal region. The constant LP buffer size aligns with the intuitive understanding that human visual processing is limited by the brain's finite bandwidth.

\subsubsection{Inverse Mapping Function}
To invert the image from the LP buffer to the screen space, we derive the inverse function using the mapping function mentioned above, known as the inverse mapping function.
\autoref{e} and \autoref{equ:y} depict the inverse transformation process of our approach, with the former depicting the calculation of the eccentric angle $e$ from the LP buffer's horizontal coordinate $u$:
\begin{equation}
	e = \frac{\exp{[(u-c_0)\cdot m/2]}-\omega_0}{m}\  ,
 \label{e}
\end{equation}
and the derivation of the screen-space position $x$, $y$ from $r$ (see \autoref{func:r}) and $v$:
\begin{equation}
	\begin{cases}
	x = r \cos {\frac{v}{l(e)}360^{\circ}} + x_0\  , \\
	y = r \sin {\frac{v}{l(e)}360^{\circ}} + y_0\  .
     \label{equ:y}
     \end{cases}
\end{equation}

\subsection{Shading Rate Adaption}
\label{subsection:segmented mapping}

\begin{figure}[tb]
	\centering 
	\includegraphics[width=0.85\columnwidth]{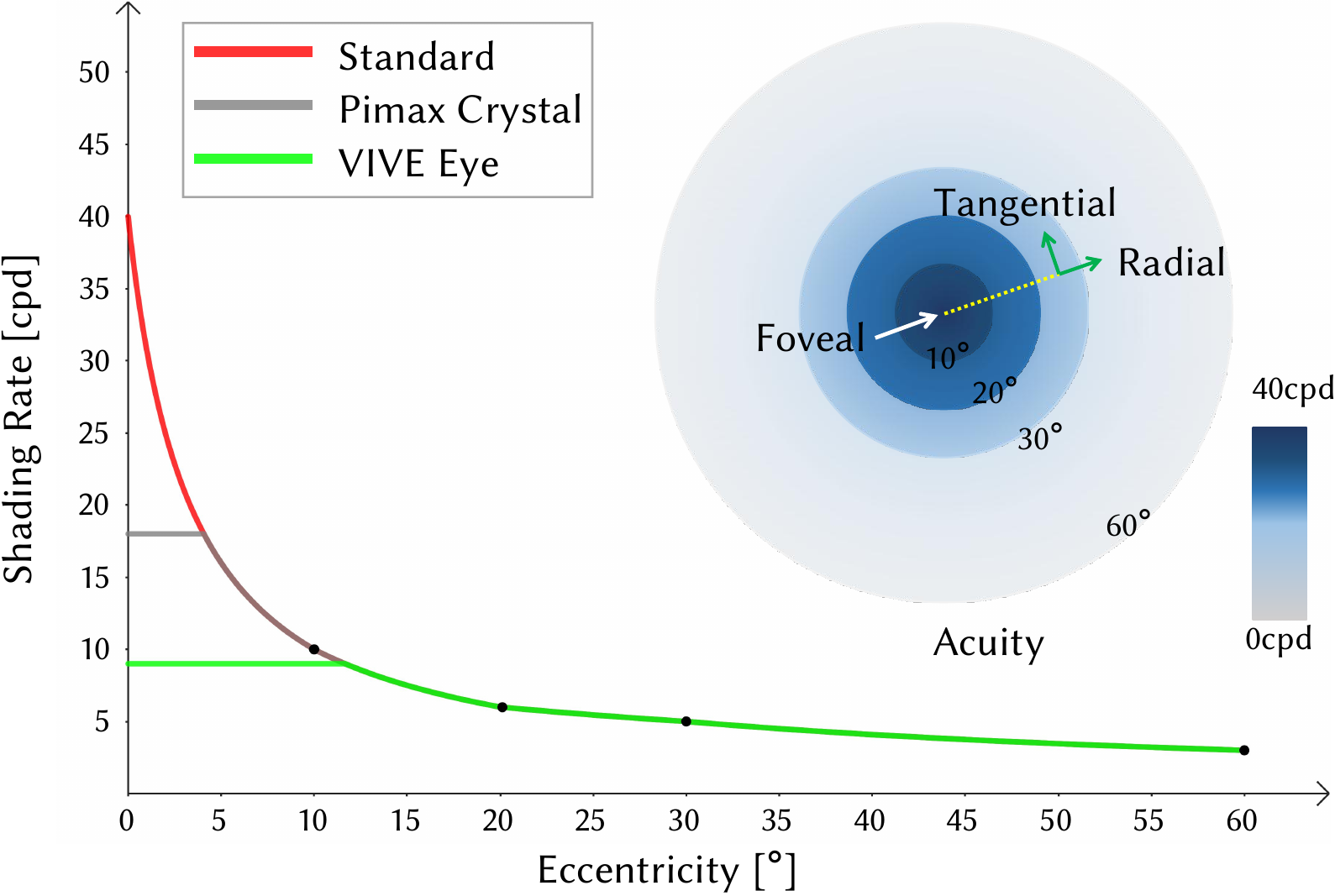}
	\caption{Our pivotal points for segmented mapping function, Standard shading rate, two curves adapted to VIVE EYE and Pimax Crystal.}
	\label{fig:ShadingRate}
\end{figure}

Examining \autoref{func:u(e)} and \autoref{func:v(e,theta)}, we find that $\omega(e)$ is a common element in both. However, recent studies \cite{Thibos:1996,Tariq:2022} suggest that visual acuity does not follow a single linear path, but is better approximated by segmented functions. Consequently, we identify specific pivot points and assign variable acuity values $f$ to precisely control the shading rate, enabling adjustments to meet the preferences of user programmers.
The detailed calculation of the shading rate and its correlation with the visual acuity will be elucidated in \Cref{semantic}.

The shading rate adaption offers a range of functions. Firstly, it enables customized shading rate control, allowing designers or users to define a tailored set of parameters for specific scenes or individual eye conditions in advance. Secondly, it supports automatic parameter tuning by detecting the VR HMD in use.
Thus, \autoref{func:u(e)} and \autoref{func:v(e,theta)} are reformulated as:
\begin{small}
\begin{equation}
	\begin{cases}
		u(e) = 2\frac{\ln (m_1e+\omega_1)}{m_1} + c_1,\  v(e) = \frac{\theta \cdot \sin{2e}}{m_1e+\omega_1},\  e\in[0,10) \\
		u(e) = 2\frac{\ln (m_2e+\omega_2)}{m_2} + c_2,\  v(e) = \frac{\theta \cdot\sin{2e}}{m_2e+\omega_2},\  e\in[10,20) \\
		u(e) = 2\frac{\ln (m_3e+\omega_3)}{m_3} + c_3,\  v(e) = \frac{\theta \cdot\sin{2e}}{m_3e+\omega_3},\  e\in[20,30) \\
		u(e) = 2\frac{\ln (m_4e+\omega_4)}{m_4} + c_4,\  v(e) = \frac{\theta \cdot\sin{2e}}{m_4e+\omega_4},\  e\in[30,60)
	\end{cases}\ .
\label{seg:u(e)}
\end{equation}
\end{small}
Regarding the inverse transformation, \autoref{e} is also divided into four segments.

In our approach, we have defined $u_{max} = u(60^{\circ})$ and $l_{max} = v(e,360^{\circ})_{max}$ as constants. For our testing purposes, we have set fixed values for $f_{0^{\circ}} = 40\ cpd$, $f_{10^{\circ}} = 10\ cpd$, $f_{20^{\circ}} = 6\ cpd$, $f_{30^{\circ}} = 5\ cpd$, $f_{60^{\circ}} = 4\ cpd$. These values are determined through a combination of data from \cite{Tariq:2022} and \cite{Thibos:1996} with a series of experimental trials. It's important to note that, at a given eccentricity, there exists a range of frequencies that are perceptible to the observer but not fully resolvable. To further enhance the perception of details, we will discuss Anti-aliasing and compatible Post-process below. Moreover, the value of these pivotal points can be increased to enhance the image quality.

The standard shading rates for our tests, based on the values above, are shown in \autoref{fig:ShadingRate}. These parameters are adjustable, allowing us to design two specific curves for the HTC VIVE EYE and Pimax Crystal headsets, each maintaining a constant foveal shading rate aligned with their maximum display densities. As shown in \autoref{fig:ShadingRate}, the HTC VIVE EYE has a lower display resolution of 9 cpd,while the Pimax Crystal offers a higher resolution of 18 cpd. Consequently, the HTC VIVE EYE curve achieves a 20\% reduction in shading load, while the Pimax Crystal curve yields a 8\% reduction.
Overall, this adaptation provides users and developers with an intuitive and convenient adjustment option.

\begin{figure}[tb]
	\centering 
	\includegraphics[width=0.95\columnwidth]{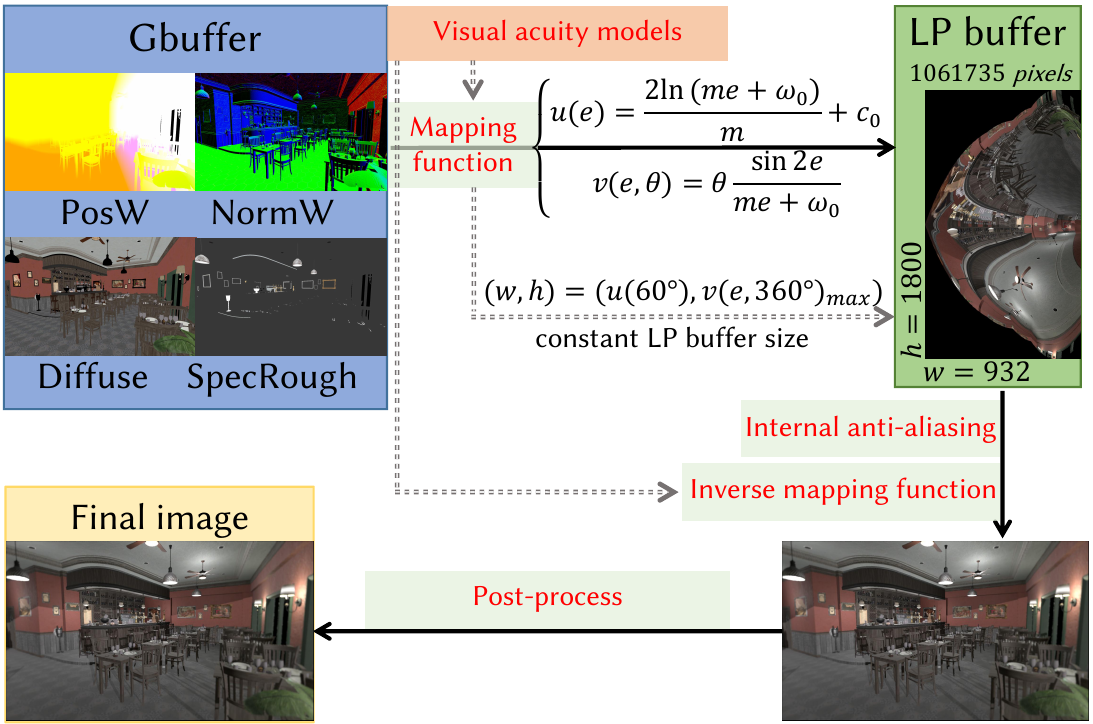}
	\caption{Working pipeline of our visual acuity consistent foveated rendering in the deferred rendering pipeline. The LP buffer dimensions are 932 $\times$ 1800, yet it includes only 1,061,735 shading points.}
	\label{fig:VaFR_framework}
\end{figure}

\subsection{Pipeline and Technical Detail}
\label{Framework}
In our pipeline (see \autoref{fig:VaFR_framework}),  the G-buffer pass runs at native resolution, with extracted data mapped to a fixed-size LP buffer, independent of display resolution. The inverse mapping function then reconstructs the screen-space image at native resolution, with internal anti-aliasing to ensure high quality. Finally, optional post-processing steps, such as temporal anti-aliasing (TAA), tone mapping, and others, can be applied to further enhance image quality. 

\textbf{Anti-aliasing:}  \ \
Given the low-resolution nature of the LP buffer, potential artifacts may occur in peripheral regions after the inverse transformation. 
We employ the Fast Approximate Anti-Aliasing (FXAA) within the LP buffer. In peripheral areas, where a few pixels in the LP buffer may cover a larger number of pixels in the screen space, precise smoothing is essential to minimize aliasing artifacts. 

LMFR \cite{Meng:2018:Kernel} uses Gaussian filters to reduce peripheral artifacts in the LP buffer, followed by foveal-aware Gaussian filters for anti-aliasing the final output. However, we found that this approach tends to blur the image without significantly enhancing visual quality. Therefore, we opt for FXAA instead.

\section{Semantic Analysis and Comparisons}
\label{semantic}
In this section, we conduct a semantic analysis and comparison of shading rate curves to identify deficiencies in prior work, along with our advantages. We first establish the methodology for calculating shading rates in log-polar-based methods in \Cref{ShadingRateAnalysis}. Using parameters from two representative studies, LMFR \cite{Meng:2018:Kernel} and LaFR \cite{Shi:2023}, we derive and visualize shading rates from their mapping functions alongside our algorithm’s shading rate. In \Cref{Inconsistency with Acuity Model}, we evaluate the alignment of each method with visual acuity. \Cref{Inconsistency under Different Resolutions} examines the adaptability of these methods to various resolutions, emphasizing the generality and extensibility of our approach.  Finally, \Cref{Inconsistency under Gaze Direction Changing} investigates the availability of each method during gaze shifts, such as saccades or smooth pursuits, in immersive VR environments. 

\subsection{Shading Rate Analysis}
\label{ShadingRateAnalysis}

From the linear model illustrated in \autoref{func:omega}, it is imperative to express the shading rate in cpd as human visual acuity model:
\begin{equation}
	SR(e) = f(e) = \frac{1}{2}\times\frac{\mathrm{d}u}{\mathrm{d}e} \ ,
\label{func:shading_equal_f}
\end{equation}
rather than assuming one percent of the unit area in LaFR~\cite{Shi:2023}. 
Here, a cycle corresponds to two pixels, where a pixel is defined as a valid shading point within the LP buffer. Consequently, the shading rate $SR(e)$ is determined by the number of cycles per unit angle at any angle $e$ within the field of view, calculated as $\frac{\mathrm{d}u}{2\mathrm{d}e}$ where $u$ represents the abscissa of the LP buffer and $e$ denotes the eccentricity angle.

The mapping equation proposed in LMFR \cite{Meng:2018:Kernel} is as:
\begin{equation}
	u_{\text{LMFR}}(z) = (z(e))^4 w\ ,
\label{func:LMFRU}
\end{equation}
and the mapping equation proposed in LaFR \cite{Shi:2023} is
\begin{small}
\begin{equation}
	u_{\text{LaFR}}(z) = \begin{cases}
		(z(e)(1-Fa))^{\frac{1}{Fa}},&e\in[0,4.89]  \\
		(\tfrac{2\arccos(Za)-\pi}{\pi})^{\tfrac{\beta}{0.7}}(1-U_{ef})+U_{ef},&e\in(4.89,55]
	\end{cases}\ ,
\label{func:LaFRU}
\end{equation}
\end{small}
where $z$ ranges from 0 to 1, representing the log ratio in the screen-space image as:
$z(e) = \frac{\ln{r}}{\ln{L}} = \frac{\ln{\frac{\tan{e}}{c_r}}}{\ln{L}}$,
and $Fa$ is short for $f'(a,\alpha)$ as:
$Fa = f'(a,\alpha) = 1-\log_{Z_{ef}}\frac{{\alpha Z_{ef}}^{1-a}}{1-a}$,
and a parameter of LaFR for its kernel function is as:
$K_f(\alpha, z(e)) = \frac{z(e)^{f'(a,\alpha)}}{1-f'(a,\alpha)}$. Differently, $a$ of LaFR is $\alpha$ of LMFR, and $\alpha$ of LaFR is a new parameter ranging from 0 to 1. Please refer to the original paper \cite{Shi:2023} for detailed information.
The $U_{ef}$ is the $U$ value at foveal boundary $e_{\text{foveal}}$ of LaFR:
$U_{ef} = (Z_{ef}(1-Fa))^{\frac{1}{Fa}}$.
The $Za$ is just a shorthand for a bunch of calculation formulas, where $Z_{ef}$ is the $Z$ value at the foveal boundary $e_{\text{foveal}}$ of LaFR:
$Za = \tfrac{\tfrac{1}{z(e)}-\tfrac{1}{Z_{ef}}}{\tfrac{1}{Z_{ef}}-1}$,
and $LW(e)$ is also a shorthand for a bunch of calculation formulas as 
$LW(e) = \frac{1}{\ln{L}}\frac{w\pi}{180\sin{e}\cos{e}}$,
where $L$ represents the farthest distance from the gaze point to the corners, and $w$ stands for their reduced-sized LP buffer width.

By taking the derivative of $e$ on both sides of the functions in both \autoref{func:LMFRU} and \autoref{func:LaFRU}, then multiplying by $0.5$, we obtain the shading rate for LMFR:
\begin{small}
\begin{equation}
	SR_{\text{LMFR}}(e) = \frac{1}{2}\times\frac{\mathrm{d}u_{\text{LMFR}}(z(e))}{\mathrm{d}e}\  = 2(\frac{\ln{\frac{\tan{e}}{c_r}}}{\ln{L}})^3 LW(e)\ ,
\label{func:LMFRSR}
\end{equation}
\end{small}
and shading rate for LaFR:
\begin{small}
\begin{equation}
	SR_{\text{LaFR}}(e) = \begin{cases}
		\frac{1-Fa}{2Fa}(z(e)-z(e)Fa)^{\frac{1-Fa}{Fa}}LW(e),&e\in[0,4.89]  \\
		\frac{2(1-U_{ef})}{-\pi\sqrt{1-Za^2}}\frac{1}{\tfrac{1}{Z_{ef}}-1}\frac{-1}{z(e)^2}LW(e),&e\in(4.89,55]
	\end{cases}\ .
\label{func:LaFRSR}
\end{equation}
\end{small}

\textbf{Our Shading Rate}:
Taking the derivative of $e$ on both sides of \autoref{func:u(e)}, the mapping function of our approach, and subsequently multiplying by $0.5$, yields the shading rate of our VaFR:
\begin{equation}
	SR_{Ours}(e) = \frac{1}{me+\omega_0} = f(e)\ ,
\label{func:OursSR}
\end{equation}
where $f(e)$ is exactly the human visual acuity model.

As our mapping function is derived from the acuity model, our approach precisely aligns the shading rate with the human visual acuity. 
Overall, we address several issues found in previous algorithms. Our approach ensures a shading rate that fully adheres to the human visual acuity model, providing users with both a comfortable experience and powerful performance.

On the contrary, LMFR and LaFR will face major issues: their shading rates do not align with human visual acuity, lack smooth adaptability to different resolutions, and cannot adjust to 3D gaze orientation changes. In the following subsections, we analyze these limitations using visual plots with different resolutions. In contrast, our algorithm will maintain a consistent shading rate that matches human visual acuity, and adapt seamlessly to resolution changes and gaze shifts common in VR. In addition, it provides user-friendly control over tangential shading rates, detailed in \Cref{Anisotropic}.

\begin{figure}[tb]
	\centering 
	\includegraphics[width=0.9\columnwidth]{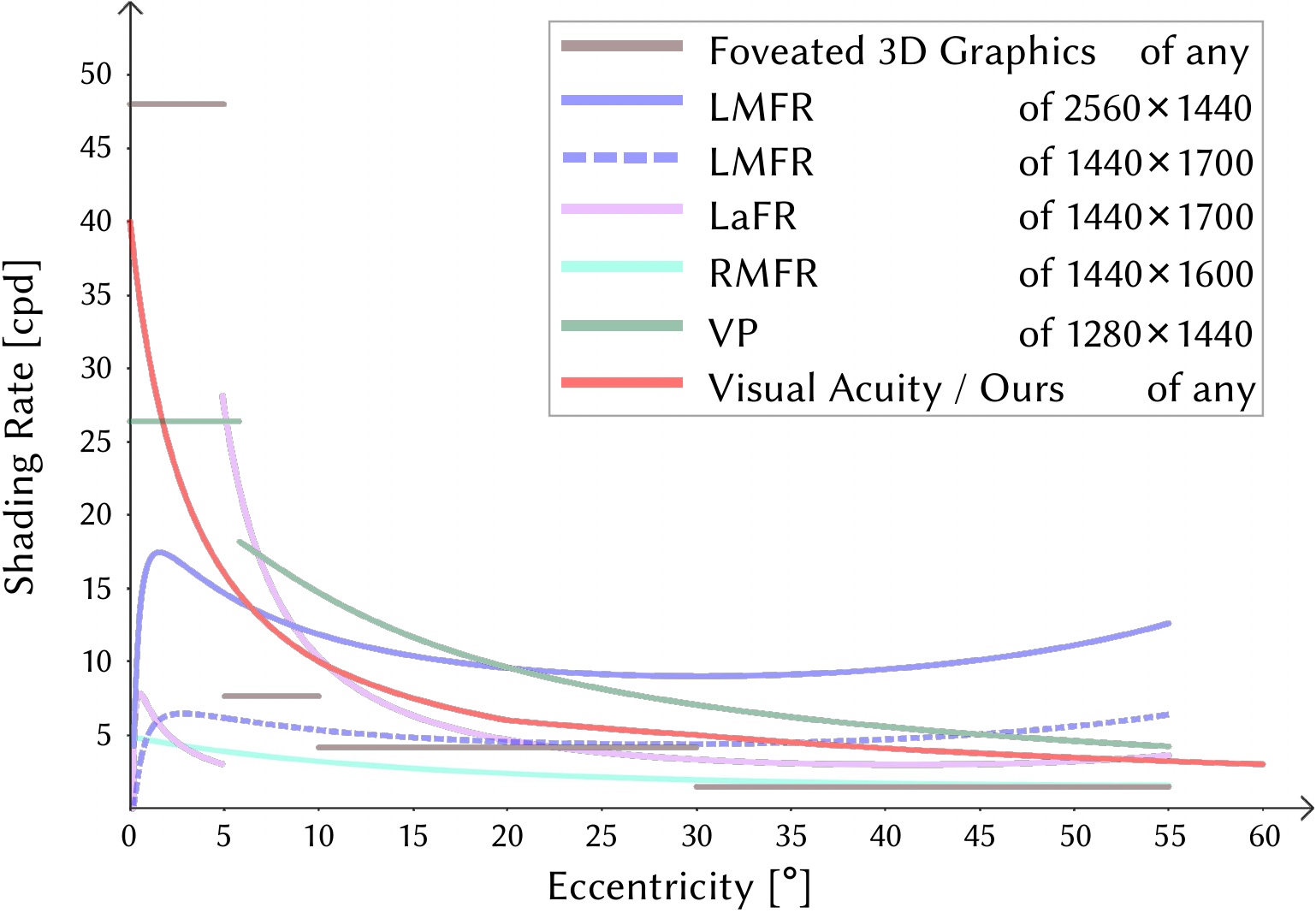}
	\caption{
    Foveated 3D graphics \cite{Guenter:2012} uses a piecewise step function, causing shading rate discontinuities. LMFR \cite{Meng:2018:Kernel} introduces a continuous shading rate but still fails to align with human visual acuity. LaFR identifies LMFR's suboptimal quality and refines it by segmenting the kernel function and fine-tuning parameters to better adapt shading distribution. RMFR \cite{Ye:2022} plots all points along the x-axis (\( f_x=0.2W, f_y=0.2H, \delta=2.6 \), gaze-centered, as per the original paper). Similarly, Visual-Polar \cite{koskela2019foveated} exhibits a shading rate far below visual acuity in the foveal region.
    }
	\label{fig:LMFRandLaFR}
\end{figure}

\begin{figure}[tb]
	\centering
	\includegraphics[width=0.9\columnwidth]{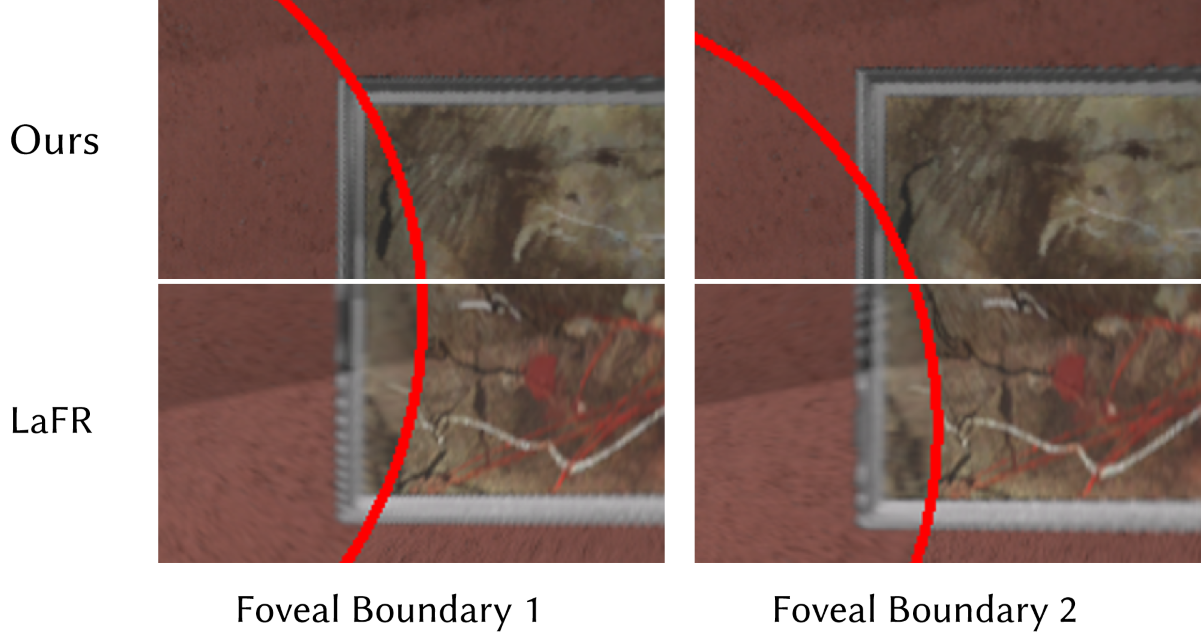}
	\caption{LaFR's shading rate discontinuity and artifacts are illustrated through ROI changes in two consecutive frames. The foveal boundary ($4.89^{\circ}$ in LaFR), marked in red, varies between left and right images. Inside this boundary, LaFR exhibits blurriness and aliasing (notably in the picture frame on the wall), while areas beyond it appear smoother. In contrast, our method eliminates these issues.}
	\label{fig:LaFRE1}
\end{figure}

\begin{figure}[tb]
	\centering
	\includegraphics[width=0.95\columnwidth]{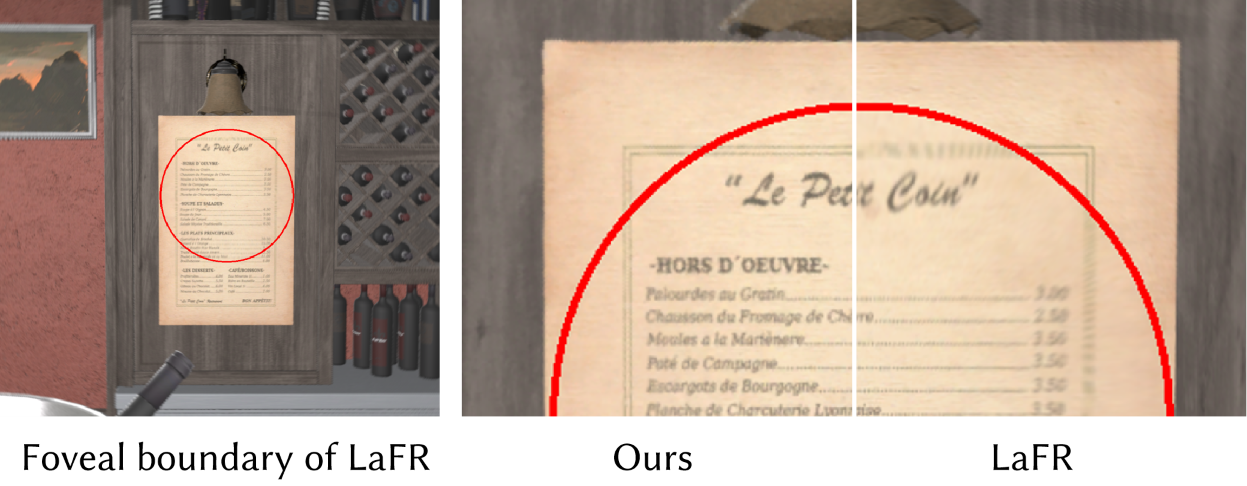}
	\caption{Side-by-side comparison of LaFR and VaFR, with the foveal boundary (defined as $4.89^{\circ}$ in LaFR) delineated in red. LaFR exhibits obvious blurriness and distortion within the foveal region, while our approach shows superior visual quality.}
	\label{fig:LaFRE2}
\end{figure}

\subsection{Evaluation of Consistency with Acuity}
\label{Inconsistency with Acuity Model}

Taking the display resolutions specified in two respective papers as examples, we conduct visual analysis of shading rates. \autoref{fig:LMFRandLaFR} showcases the shading rate of foveated 3D graphics by Guenter et al. \cite{Guenter:2012} and LMFR, utilizing the resolution ($2560\times1440$) defined in LMFR. Switching to another resolution ($1440\times1700$) defined in LaFR, \autoref{fig:LMFRandLaFR} depicts the shading rate of several comparable algorithms, exposing issues associated with both LMFR and LaFR.

\textbf{Guenter et al. \cite{Guenter:2012}}: Typical multi-resolution rendering involves partitioning the field of view and rendering at different resolutions, followed by synthesizing the final image. Coarse Pixel Shading (CPS) \cite{Vaidyanathan:2014} introduces coarse-grained pixel rendering to share shading values among pixels within a defined range. Methods based on Variable Rate Shading (VRS), akin to CPS, transfer coarse-grained partitioning from software to hardware. The shading rates generated by these methods are discrete, roughly illustrated in \autoref{fig:LMFRandLaFR}. Clear boundaries between sharp and blurred blocks significantly impact the user's sensory experience in VR.

\textbf{LMFR \cite{Meng:2018:Kernel}}: LMFR approximates human visual acuity in specific resolution settings, with its fitting being more accurate within the 3°-15° range. However, beyond this range, the shading rate in the peripheral field exceeds human visual acuity, resulting in significant computational resource waste. Furthermore, the manually configured kernel function used to redistribute shading points limits the shape of the shading rate curve, narrowing its fitting range.

\textbf{LaFR~\cite{Shi:2023}}:
At LMFR's specified resolution, its shading rate closely approximates the visual acuity model. However, this alignment breaks down when switching to the resolution used in LaFR. LaFR segments the kernel function and introduces new parameters to adjust the shading points, closely approximating the human visual acuity within $8^{\circ}-55^{\circ}$ (see \autoref{fig:LMFRandLaFR}). However, truncating the shading rate results in discontinuities that degrade image quality in the foveal region, and impact user experience, as evidenced by our experiment (see \autoref{fig:LaFRE1} and \autoref{fig:LaFRE2}).

LMFR’s shading rate parameters are not adaptable to different resolutions, leading to misalignment with the acuity curve. For both LMFR and LaFR, this reveals the limited universality of the parameters derived from user studies. Additionally, LaFR requires fine-tuning, such as the kernel function segmentation and new parameters, to recalibrate shading distribution for new HMD resolutions. This need for case-by-case adjustments highlights the complexity of achieving optimal shading rates across varying resolutions.

\textbf{RMFR~\cite{Ye:2022}}:
Rectangular mapping foveated rendering (RMFR) also involves transformations between two Cartesian coordinate systems; however, our shading rate calculation in \autoref{func:shading_equal_f} based on the log-polar approach, no longer applies for capturing shading rates in all directions for RMFR. Therefore, we visualize only the radial and tangential shading rates along the x-axis, using ratios relative to window height and width as defined in the original paper. At RMFR's resolution of $1440\times1600$ with a $\delta$ of $2.6$, the shading rates fall significantly below the human visual acuity curve, especially in the tangential direction (see \autoref{fig:OurRadialTangential}).
Since rectangular mapping follows a different technical approach and has distinct features from ours, we will not include comparisons with RMFR in the following sections.

\textbf{VP~\cite{koskela2019foveated}}:
Visual-Polar method introduces a log-polar approach with linear fall-off in the foveal region, performing partial rendering of the LP buffer to create a cropped triangular area, ensuring uniform shading rates within the foveal region. Outside of this area, the method employs a different visual acuity model and a gaze-related parameter $d$, which can lead to inconsistencies during saccades. Interestingly, the LP buffer size used here is similar to ours, resulting in comparable shading rate curves in the peripheral region. However, it shows a shading rate that falls far below visual acuity in the foveal region. In addition, this approach still does not decouple the radial and tangential shading rates, causing the tangential shading rate to depend on the radial rate and to be lower.

\begin{figure}[tb]
	\centering 
	\includegraphics[width=0.9\columnwidth]{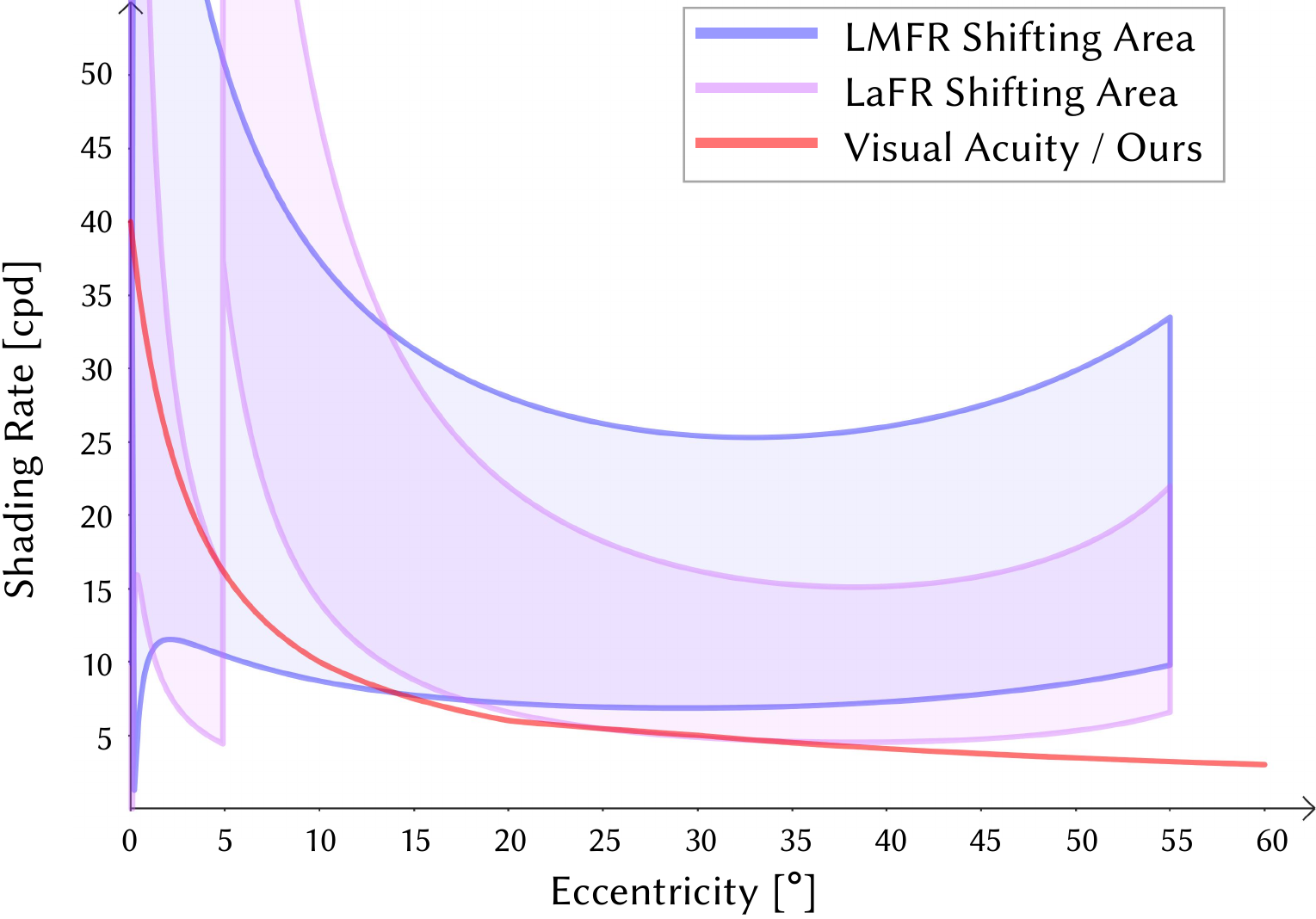}
	\caption{As the resolution increases from $1920\times1080$ to $7680\times4320$, the shading rates of LMFR and LaFR deviate significantly from human acuity, indicating resolution dependence. In contrast, our method consistently aligns with the acuity curve across resolutions, showcasing its adaptability and robust capabilities.
}
	\label{fig:ResolutionChange}
\end{figure}

\subsection{Adaptability Across Different Resolution}
\label{Inconsistency under Different Resolutions}

The shading rates of LMFR and LaFR are sensitive to changes in resolution due to the term $w = \frac{W}{\delta}$ in their shading rate functions, where $W$ represents the width of the display resolution and $\delta$ is the reduction ratio (see details in \Cref{subsec:foveated}).
As shown in \autoref{fig:ResolutionChange}, transitioning from $1920\times1080$ to $7680\times4320$ results in significant changes in the shading rate, represented by the colored shifting area. This indicates that LMFR and LaFR are optimized for specific resolutions, making them unsuitable for use on different devices where their parameters may lose effectiveness. The requirement of using a kernel function $K(\cdot)$ to redistribute shading points to fit human visual acuity presents inherent challenges. Mapping all display resolution pixels into a proportionally reduced shading buffer is closely tied to the display resolution and reduction ratio, despite efforts to fine-tune parameters through user experiments.

\begin{figure}[tb]
	\centering 
	\includegraphics[width=0.9\columnwidth]{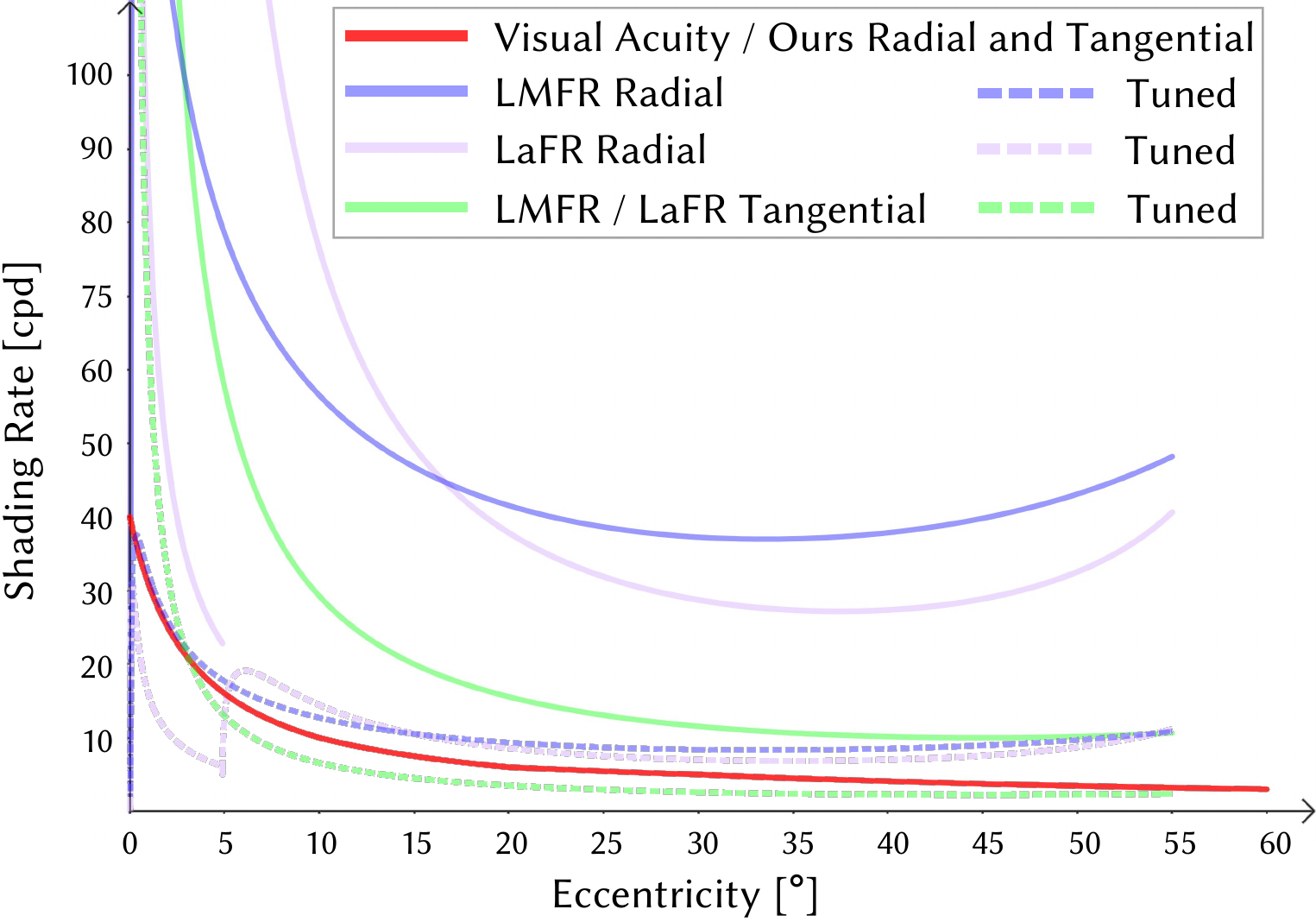}
	\caption{At a retinal resolution of $11520\times6480$, prior methods inefficiently use computing resources, as their shading rate unnecessarily exceeds the actual requirements of the human eye. We tuned the parameters for LMFR and LaFR, setting $\delta=8$, $\alpha=0.78$, and $\beta=0.9$, in an attempt to optimize their shading rate closer to human visual acuity, though they can hardly achieve this alignment.
    }
	\label{fig:retinal}
\end{figure}

Assuming a visual field of $100^\circ$ vertically and $120^\circ$ horizontally, with a foveal acuity of 40 cpd (where each cycle equals two pixels), the theoretical retinal resolution amounts to an astonishing $100\times80\times120\times80 = 76,800,000$ pixels per eye. For practical purposes, we approximate this as $11520\times6480 = 74,649,600$ pixels per eye.
As illustrated in \autoref{fig:retinal}, at a retinal-level resolution of $11520\times6480$, LMFR and LaFR exhibit markedly elevated shading rates, surpassing practical requirements and resulting in inefficient resource utilization.

In contrast, our method consistently conforms to the visual acuity and remains unaffected by changes in resolution. We avoid the need for kernel functions and instead derive our approach directly from the acuity model. This involves calculating the information density perceivable by the human eye in a single frame, allowing us to establish a fixed LP buffer size that is directly correlated with visual acuity. This fixed buffer size remains unchanged regardless of device or resolution variations, as the spatial information density perceptible to the human eye remains constant.

\begin{figure}[tb]
	\centering 
	\includegraphics[width=0.9\columnwidth]{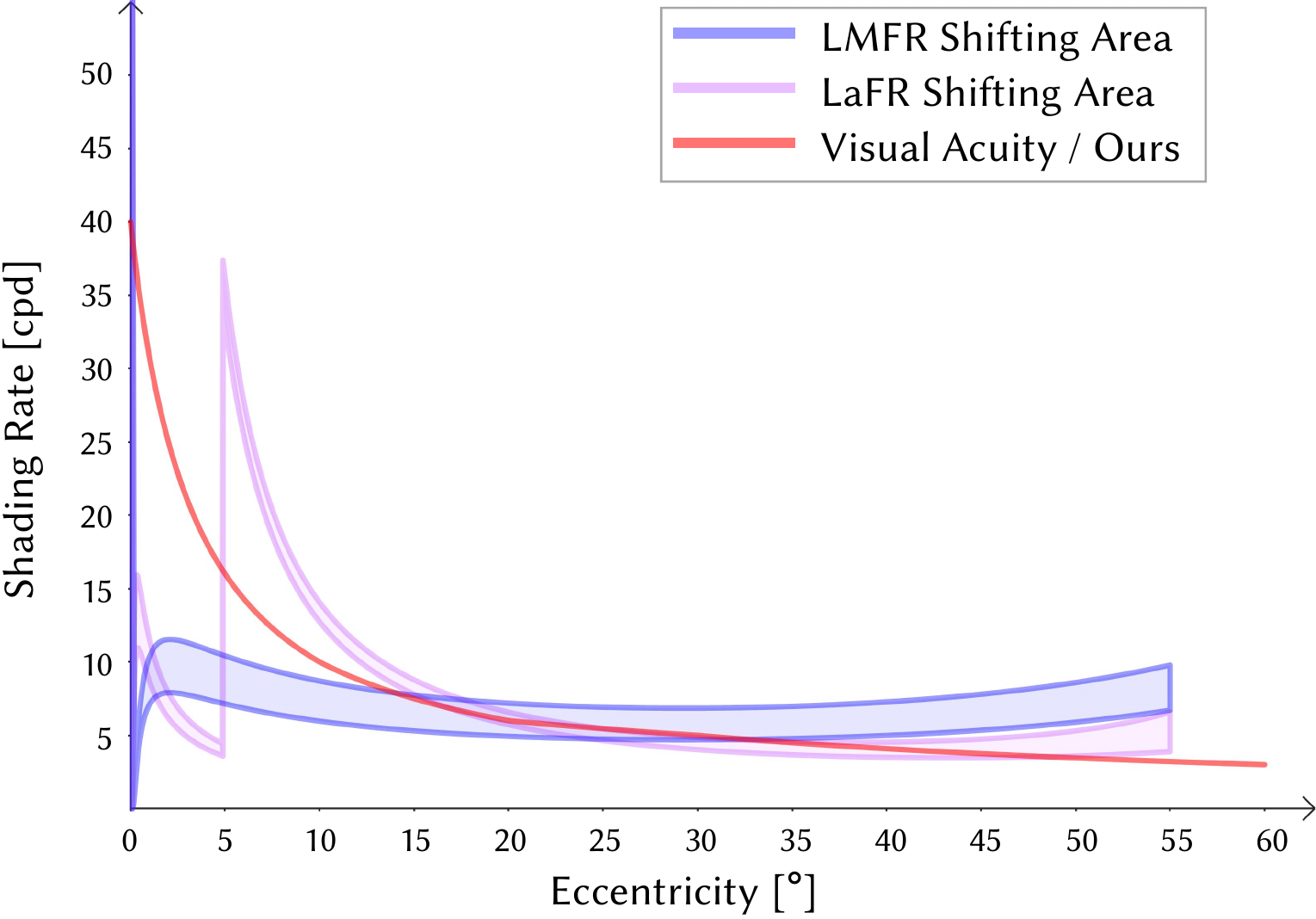}
	\caption{Shading rates of LMFR and LaFR vary during gaze transfer from looking straight ahead to another direction (gaze deviation), resulting in perceptual inconsistencies. In contrast, our approach maintains a consistent alignment with visual acuity, ensuring stability regardless of gaze shift.
    }
	\label{fig:GazeChange}
\end{figure}

\subsection{Shading Rate During Gaze Shift}
\label{Inconsistency under Gaze Direction Changing}
Typically, an individual's gaze point dynamically shifts, naturally transitioning to different regions of interest. This gaze transfer behavior is pivotal in comprehending how our visual perception adjusts to diverse stimuli and environments. Moreover, variations in gaze orientation significantly affect tasks involving interaction with digital interfaces.

As shown in \autoref{fig:GazeChange}, illustrating the rendering at a resolution of $2560\times1440$ in VR, a noticeable shift occurs in the shading rate when the gaze point shifts from the center to the corner of the screen. The colored shifting area delineates the variation in shading rate, illustrating the potential for sensory discontinuity. This phenomenon, where image quality varies with eye movement, arises from the use of $ln(L)$ in the denominators of their log-polar transformations, as indicated in \autoref{func:LMFRU} and \autoref{func:LaFRU}, inherently correlating with gaze orientation.

In contrast, our method deviates from the traditional log-polar transformation used by LMFR and LaFR. By exclusively harnessing the human visual acuity model, our algorithm maintains independence from the gaze orientation. This ensures uniformity in the VR experience with a stable shading rate across different gaze orientations.

\subsection{Parameters Tuning}

In addition to the parameter values $\delta=1.8$, $\alpha=0.85$, and $\beta=0.7$ used in the original paper, we further tuned the parameters for LMFR and LaFR, setting $\delta=8$, $\alpha=0.78$, and $\beta=0.9$ to allow for more rigorous comparisons. These values aim to optimize the shading rates of LMFR and LaFR to closer align with human visual acuity at retinal resolution.

As shown in \autoref{fig:retinal}, although parameter tuning allows the LMFR and LaFR methods to reduce the number of shading pixels for practical use at extremely high resolutions, their lack of generalizability remains a drawback. Furthermore, parameter tuning cannot alter the curve shape dictated by the mapping equations, meaning that these curves can only approximate, rather than perfectly fit, the human visual acuity model. Lastly, the tangential shading rate curves produced by these methods remain suboptimal.

In contrast, our approach, without any parameter tuning, maintains a consistent shading rate aligned with human visual acuity, enabling retinal-level detail in the foveal region while providing satisfactory quality in peripheral areas.
Furthermore, VaFR's radial and tangential shading rates are identical because of our setting in \autoref{func:tangential}. Further details on the tangential shading rate are discussed in \Cref{Anisotropic}. Even at retinal resolutions where conventional methods falter, our approach remains robust. Tailored for forthcoming high-end VR headsets boasting retinal-level displays, our methodology aspires not only to sustain a seamless frame rate but also to elevate image fidelity to the retinal level, thus advancing the frontiers of immersive experiences in virtual reality.

\section{Validation on Deferred Rendering Pipeline}

To evaluate both the rendering quality and performance of our method, we have designed a comprehensive testing framework using C++ with the NVIDIA Falcor framework \cite{benty2020falcor}. Our setup includes a PC workstation equipped with a 3.2 GHz Intel(R) Core(TM) i7-8700 CPU, 8 GB of memory, and an NVIDIA GeForce GTX 2080 SUPER graphics card.

\subsection{Integration with Reprojection}

Extensive experiments have shown that rasterization is the primary bottleneck in high-resolution deferred rendering, especially in stereo rendering. Reprojection for binocular VR can reduce the rasterization workload, increasing performance when integrated into our deferred rendering pipeline.

To test the upper limit and compatibility of our VaFR, we integrate it with Wissmann et al.'s reprojection strategy \cite{Wissmann:2020}, creating VaFR$_{\text{Rep}}$. This approach reprojects left-eye images to the right perspective using forward grid warping. Unlike Wissmann’s method, we perform reprojection directly from the left log-polar buffer, reducing computational load and significantly enhancing performance. In contrast, the non-reprojection counterpart, VaFR$_\text{{Dup}}$, renders the left- and right-eye images independently using the VaFR method, performing separate renderings twice. Detailed results are presented in \Cref{Sec:Performance Evaluation} and \Cref{Sec:user study}, with additional benefits discussed in \Cref{limitation}.

\subsection{Performance Evaluation}
\label{Sec:Performance Evaluation}
\subsubsection{Setup and Procedure}
\label{setupandprocedure}
Experiments are conducted using our VaFR series methods as the experimental group and several existing algorithms as the control group, including LaFR(2.2), LaFR(1.8), LMFR(1.8), and GT. In this context, GT refers to the native rendering algorithm used as the baseline, identical to the other comparable algorithms but without foveated rendering acceleration. According to the results of X. Shi \cite{Shi:2023}, LaFR achieves different rendering ratios by adjusting the parameter $\delta$ within the range of 1.8 to 2.2, which means that the shading resolution equals the display resolution divided by ${\delta^2}$. LaFR(1.8) achieves better rendering quality, while LaFR(2.2) achieves higher rendering performance. 
All parameters of LMFR and LaFR are used exactly as stated in the original papers. Anti-aliasing, as specified in LMFR \cite{Meng:2018:Kernel}, is implemented accordingly. Since LaFR does not mention anti-aliasing, we assume that it uses an approach analogous to that of LMFR.

NVIDIA's Bistro and Sun Temple are used for the testing scenes. In BistroExterior, there is one directional light for the daylight scene, 12 light sources for the nighttime scene, 21 lights for the BistroInterior scene, and 31 lights for the Temple scene. Note that the complexity of the scene and the number of lights can heavily affect the rendering workload.
During the rendering process, performance metrics are recorded for all approaches, encompassing VaFR$_\text{{Rep}}$, VaFR$_\text{{Dup}}$, LaFR(2.2), LaFR(1.8), LMFR(1.8), GT. The performance is evaluated by the averaged frame rate across various resolutions, from 2K per eye to retinal resolution.

We begin our analysis with the BistroExteriorNight scene, which features 12 lights, followed by the BistroExteriorDay scene, illuminated by a single directional light source, representing scenarios with minimal lighting complexity. To further investigate the performance, we use the BistroInterior scene with 21 lights, which has one-third of the triangles and twice lights compared to BistroExteriorNight. This setup allows us to explore the impact of reducing triangle count and increasing light sources on rasterization and lighting loads. Furthermore, given that VaFR$_\text{{Dup}}$, LMFR, and LaFR aim to enhance performance by reducing shading computations, we extend our tests to the Temple scene, which has fewer triangles than BistroInterior but features 31 lights.

\begin{figure}[tb]
	\centering 
	\includegraphics[width=\columnwidth]{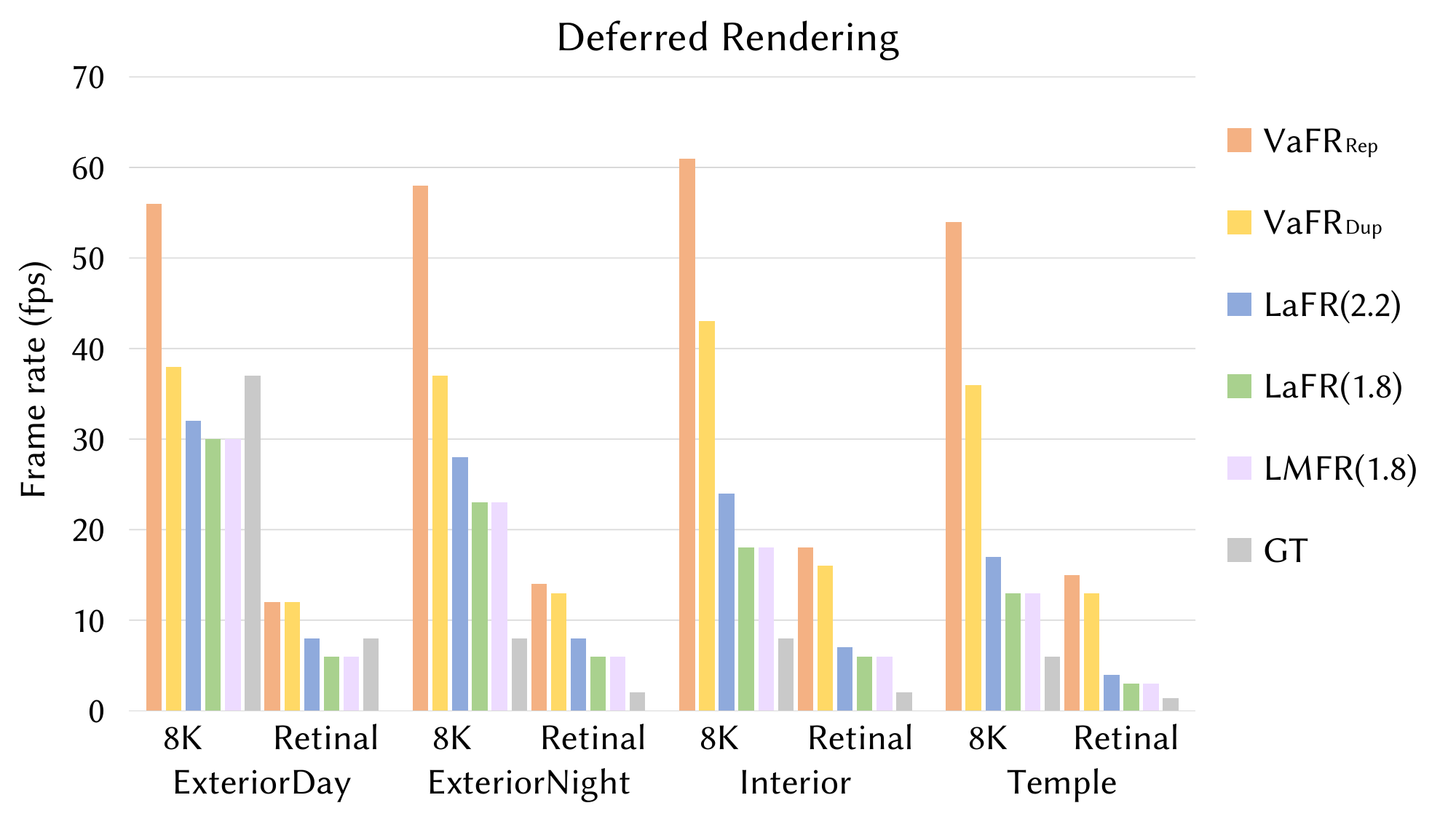}
	\caption{Performance comparisons of different approaches in various scenes, 8K and retinal resolutions are evaluated, respectively. VaFR$_\text{{Rep}}$ achieves over 54 fps at 8K, whereas other modes fall below 30 fps.}
	\label{fig:PerformanceDR}
\end{figure}

\begin{figure}[tb]
	\centering 
	\includegraphics[width=\columnwidth]{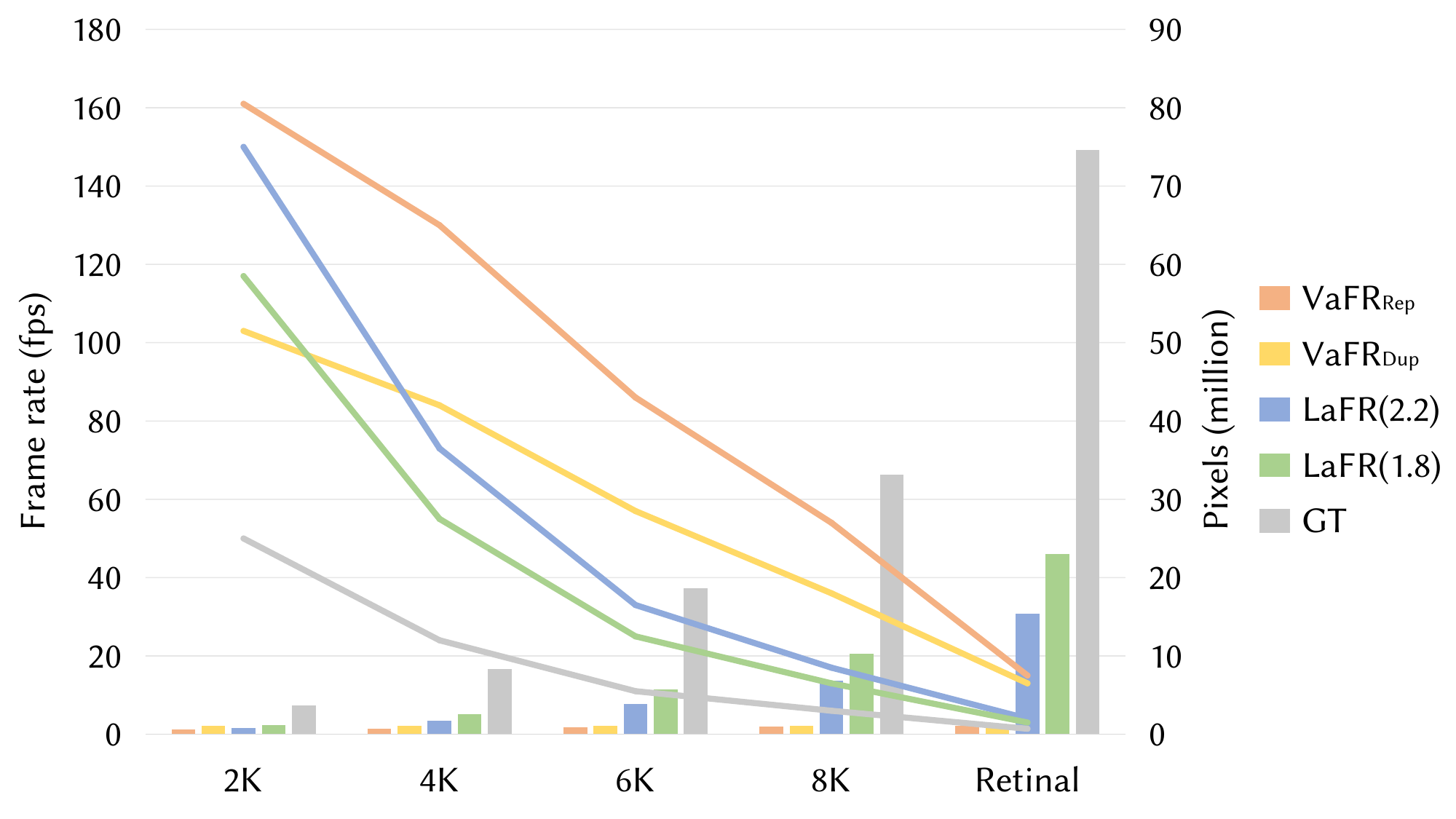}
	\caption{Performance of deferred rendering declines when the resolution is increased from 2K to retinal level, using the Temple scene for testing. The histogram illustrates the number of pixels rendered per eye.}
	\label{fig:DecreaseCurveDR}
\end{figure}

\begin{table}[tb]  
\caption{Number of pixels/rays of each rendering mode.}
\begin{center}     
\setlength\tabcolsep{2pt}
\begin{tabular}{c|c|c|c|c|c}   
\hline    Quantity                     & VaFR$_\text{{Rep}}$ & VaFR$_\text{{Dup}}$ & LaFR(2.2) & LaFR(1.8) & GT   \\ 
\hline   2K  & 637K & 1,061K  & 761K  & 1,137K &3,686K \\  
\hline   4K  & 743K & 1,061K  & 1,713K &2,560K & 8,294K \\ 
\hline   6K  & 849K & 1,061K  & 3,855K &5,760K & 18,662K\\ 
\hline   8K  & 955K & 1,061K  & 6,854K &10,240K & 33,177K \\ 
\hline   Retinal &1,035K& 1,061K&15,423K&23,040K & 74,649K \\ 
\hline   
\end{tabular}   
\end{center}
\label{table:pixels or rays} 
\end{table}

\subsubsection{Result}
\label{result}

Comparing BistroExteriorDay, which has one directional light, it is evident that LMFR, LaFR, and VaFR$_\text{{Dup}}$ all lead to a decrease in performance. This is because the lighting pass has minimal influence on overall performance, due to only one light in the scene, with the main bottleneck arising from rasterization. However, VaFR$_\text{{Rep}}$ achieves speedup owing to the elimination of rasterization processing of the right-eye image. Detailed performance is shown in \autoref{fig:PerformanceDR}.

In contrast, BistroInterior presents more lights and fewer triangles. As resolution increases, LMFR and LaFR maintain a relatively constant speedup of 2$\times$, while VaFR$_\text{{Dup}}$ demonstrates a much higher speedup. At the highest resolution, VaFR$_\text{{Dup}}$ achieves a $8\times$ speedup and VaFR$_\text{{Rep}}$ attains a $9\times$ speedup.

In the most challenging scenario, involving the Temple scene with 31 lights and 606,376 triangles, LaFR(2.2) achieves a speedup of 2.9$\times$ to 3$\times$, while LaFR(1.8) and LMFR(1.8) achieve a speedup of 2.1$\times$ to 2.3$\times$. VaFR$_\text{{Dup}}$ exhibits a remarkable $9.3\times$ speedup and VaFR$_\text{{Rep}}$ demonstrates an outstanding $10.7\times$ speedup.

However, due to rasterization demands in the deferred rendering, which require generating G-buffers at the display resolution and creating several large textures for the lighting pass, this process represents a significant portion of the rendering time. As a result, methods based on the deferred rendering often show relatively low efficiency, particularly at extremely high resolutions, such as 8K or retinal levels. 
Notably, as the resolution exceeds 4K, LaFR loses its performance advantage and undergoes a marked decline. 

Since LaFR(1.8) and LMFR(1.8) generally exhibit similar frame rates, we only show LaFR(1.8)'s performance as a representative for simplicity. \autoref{table:pixels or rays} lists the number of pixels rendered per eye of each rendering mode across different resolutions.
\autoref{fig:DecreaseCurveDR} illustrates this drop in frame rate when the resolution increases from 2K to the retinal level in the Temple scene. In contrast, the ray casting pipeline, not requiring the rasterization of very large resolution G-buffers, proves to be more suitable and performant at these high resolutions, as we will discuss in \Cref{sec:ray casting}.

\begin{figure}[tb]
	\centering 
	\includegraphics[width=0.95\linewidth]{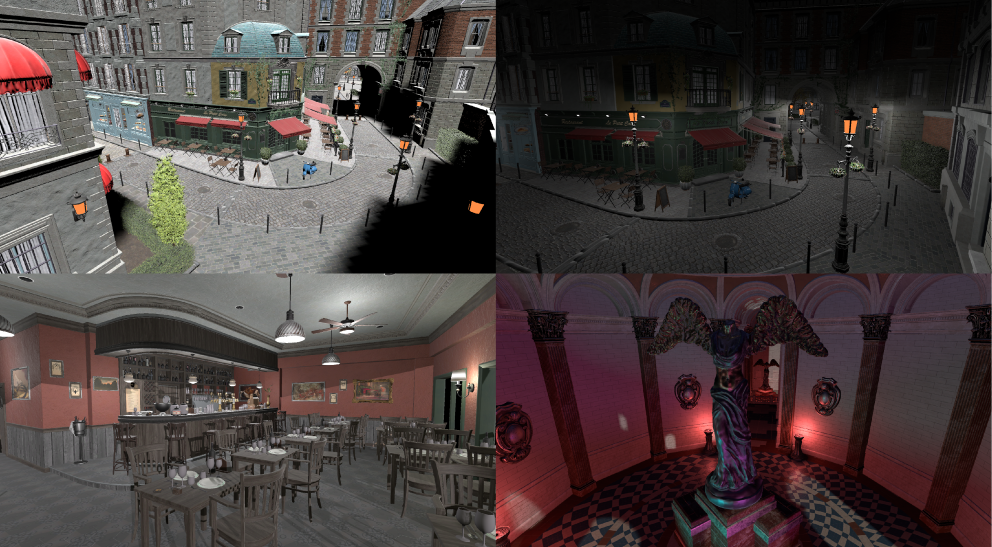}
	\caption{Four benchmark scenes are employed for testing, including BistroExteriorDay, BistroExteriorNight, BistroInterior, and SunTemple. }
	\label{fig:scenes}
\end{figure}

\subsection{User Study}
\label{Sec:user study}
\subsubsection{Pilot User Study}
We conducted a user study, testing it in an immersive virtual reality environment equipped with various types of head-mounted displays (HMDs) across different scenarios. This is done to compare real users' preferences for the experience provided by our method against alternative algorithms.

\textbf{Setup}  \ \ 
An HTC VIVE PRO EYE is used, with a built-in gaze tracking system, to precisely monitor user's gaze orientation. This HMD features a display resolution of $1440\times1600$ pixels per eye, while the actual rendered texture resolution (also called rendering resolution) is $2468\times2740$ pixels per eye. Additionally, a Pimax Crystal is used, also equipped with a gaze tracking system, boasting a much higher display resolution of $2880\times2880$ pixels per eye, achieving up to 35 pixels per degree, with the actual rendered texture resolution of $3108\times2624$ pixels per eye.

\textbf{Participant}  \ \
20 participants (17 males and 3 females, aged 22 to 28) were recruited, all of whom were graduate students at our university with normal or corrected-to-normal vision.

\textbf{Procedure}  \ \
Each participant's session lasted approximately 25 to 30 minutes, involving free exploration of three scenes in five modes from different approaches. To minimize bias, we avoided providing information about our research or the rendering algorithm. Once participants comfortably wear the headset, they will be exposed to various rendering approaches in random order and select their top two preferences. They could choose only a single preference if they could not decide on two. Additionally, after evaluating each approach, participants could revisit specific rendering methods before finalizing their decision. Subsequently, they moved on to assess the next scene for testing.
Two HMDs listed above were used for our user study.

\begin{figure}[tb]
	\centering 
	\includegraphics[width=0.95\columnwidth]{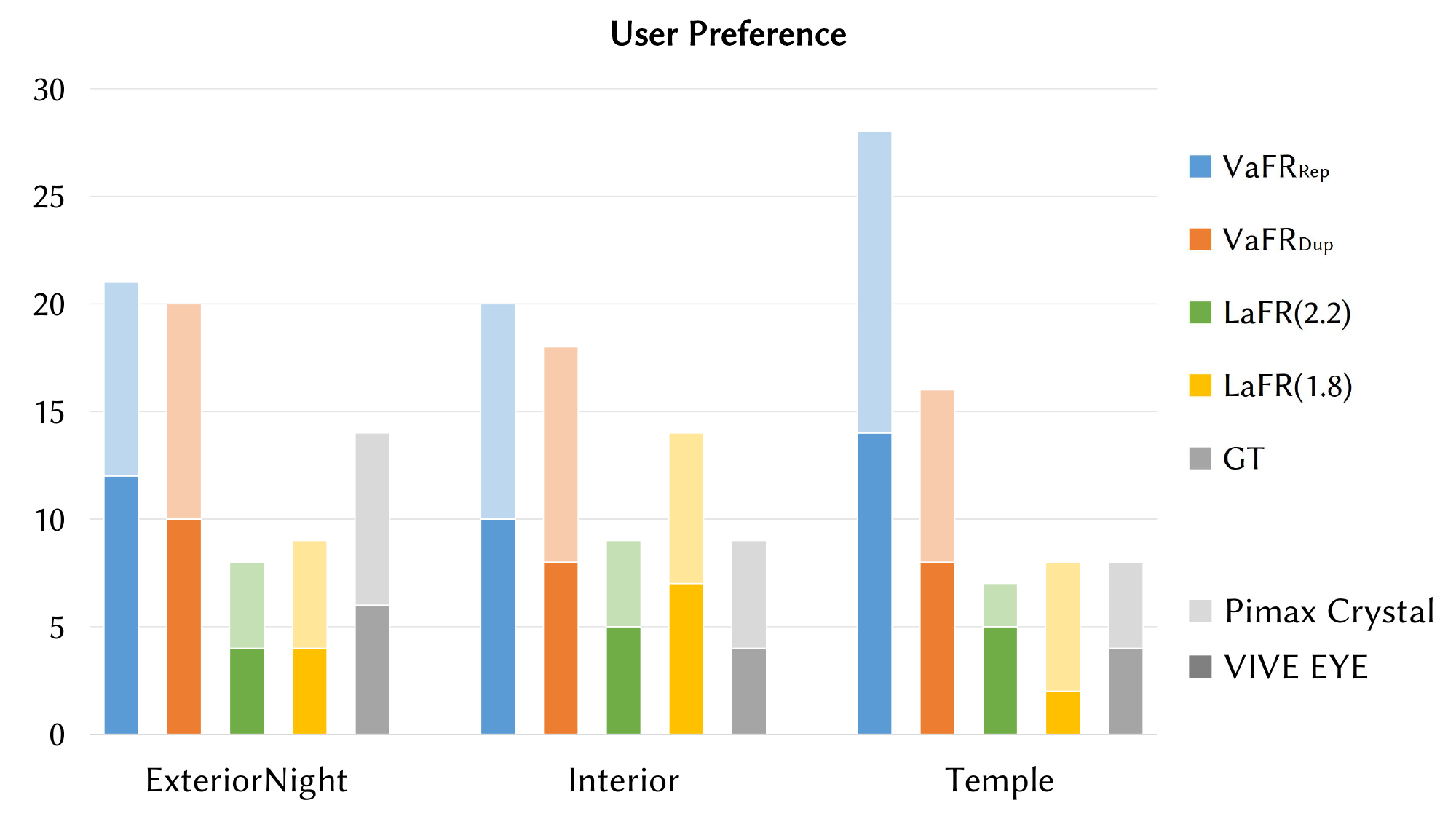}
	\caption{User preference using VIVE EYE HMD with a rendering resolution of $2468\times2740$ per eye and Pimax Crystal headset with a rendering resolution of $3108\times2624$ per eye, VaFR$_\text{{Rep}}$ and VaFR$_\text{{Dup}}$ receive the most votes.}
	\label{fig:Preference2in5}
\end{figure}

\begin{table}[tb]  
\caption{Kruskal-Wallis test for pairwise comparisons with Holm-Bonferroni correction. At a significance level of 0.05, both VaFR$_\text{{Rep}}$ and VaFR$_\text{{Dup}}$ show p values smaller than 0.05, signifying that our approaches are significantly superior to the alternatives.}
\begin{center}     
\setlength\tabcolsep{2pt}
\begin{tabular}{c|c|c|c|c|c}   
\hline    P values                     & VaFR$_\text{{Rep}}$ & VaFR$_\text{{Dup}}$ & LaFR(2.2) & LaFR(1.8) & GT   \\ 
\hline   VaFR$_\text{{Rep}}$ &    /                        &0.177  &  0.028   & 0.023    & 0.025 \\  
\hline   VaFR$_\text{{Dup}}$                        &    /                        &  /   &  0.031   & 0.031    & 0.032 \\ 
\hline   LaFR(2.2)                     &    /                        &  /   &   /     & 0.562    & 0.525\\ 
\hline   LaFR(1.8)                     &    /                        &  /   &   /     &   /     & 0.806 \\ 
\hline   
\end{tabular}   
\end{center}
\label{table:Kruskal-Wallis2in5} 
\end{table}

\subsubsection{Results and Analysis}
\label{userAnalysis-P}
\autoref{fig:Preference2in5} illustrates the user preferences observed in the VIVE EYE and Pimax Crystal HMDs.
Evidently, VaFR$_\text{{Dup}}$ and VaFR$_\text{{Rep}}$ consistently received top user preferences across scenes. However, users favored GT in the BistroExteriorNight scene for its smooth frame rate and superior image quality. Conversely, in the Temple scene with complex lighting, performance of LaFR and GT was notably affected, whereas VaFR$_\text{{Rep}}$ and VaFR$_\text{{Dup}}$ perform better in handling such conditions, further highlighting our algorithm's capability in challenging scenarios. 

In addition, we established null hypotheses for the user study: There is no significant difference in user preference data for specific rendering modes between any two groups. As indicated in \autoref{table:Kruskal-Wallis2in5}, the Kruskal-Wallis test was applied for pairwise comparisons at a significance level of 0.05, with adjustments using the Holm-Bonferroni correction. The corrected P values for VaFR$\text{Rep}$ and VaFR$\text{Dup}$ are less than 0.05, signaling a significant difference compared to other methods. This suggests a violation of the null hypothesis, indicating that our approach outperforms alternative approaches. In contrast, the P values for LaFR(2.2), LaFR(1.8), and GT following the Kruskal-Wallis test exceed 0.05, implying no significant differences among them.

\subsubsection{2AFC User Study}
To further validate the findings of the pilot user study, we conducted an additional two-alternative forced choice (2AFC) test.

\textbf{Participant}  \ \
25 participants (17 males and 8 females, aged 22 to 32, with a mean age of 25.56) were recruited, primarily graduate students from our university, all of whom were graduate students from our university with normal or corrected-to-normal vision. Some participants in the pilot study were included again, and the setup remained the same as in the pilot study.

\textbf{Procedure}  \ \
In each trial of 2AFC test, they are exposed to two rendering approaches and select only one as their preference. Participants could revisit any rendering methods before finalizing their decision. Each pair of rendering modes is tested, resulting in a total of 10 combinations. Each combination presents 3 times, with the order within each combination being randomized, as well as the order between combinations. Thus, each scene has 10 (combinations) $\times$ 3 (times)= 30 trials.
After completing each set, the participants proceed to the next scene for testing. Two HMDs listed above were employed in the user study. In total, 3 (scenes) $\times$ 30 (trials per scene) $\times$ 25 (participants) $\times$ 2 (HMDs)= 4500 trials were conducted.

Regarding preferences, participants were informed in advance to base their preferences solely on their subjective visual experience while navigating the 3D scenes, without any specific guidelines. After completing the experiment, each participant was asked to explain their choices. Most preferred a balanced experience, valuing both frame rate and image quality. A minority, potentially more susceptible to 3D motion sickness, prioritized frame rate, while only two participants highlighted the importance of image quality.

\begin{figure}[t]
	\centering 
	\includegraphics[width=0.85\columnwidth]{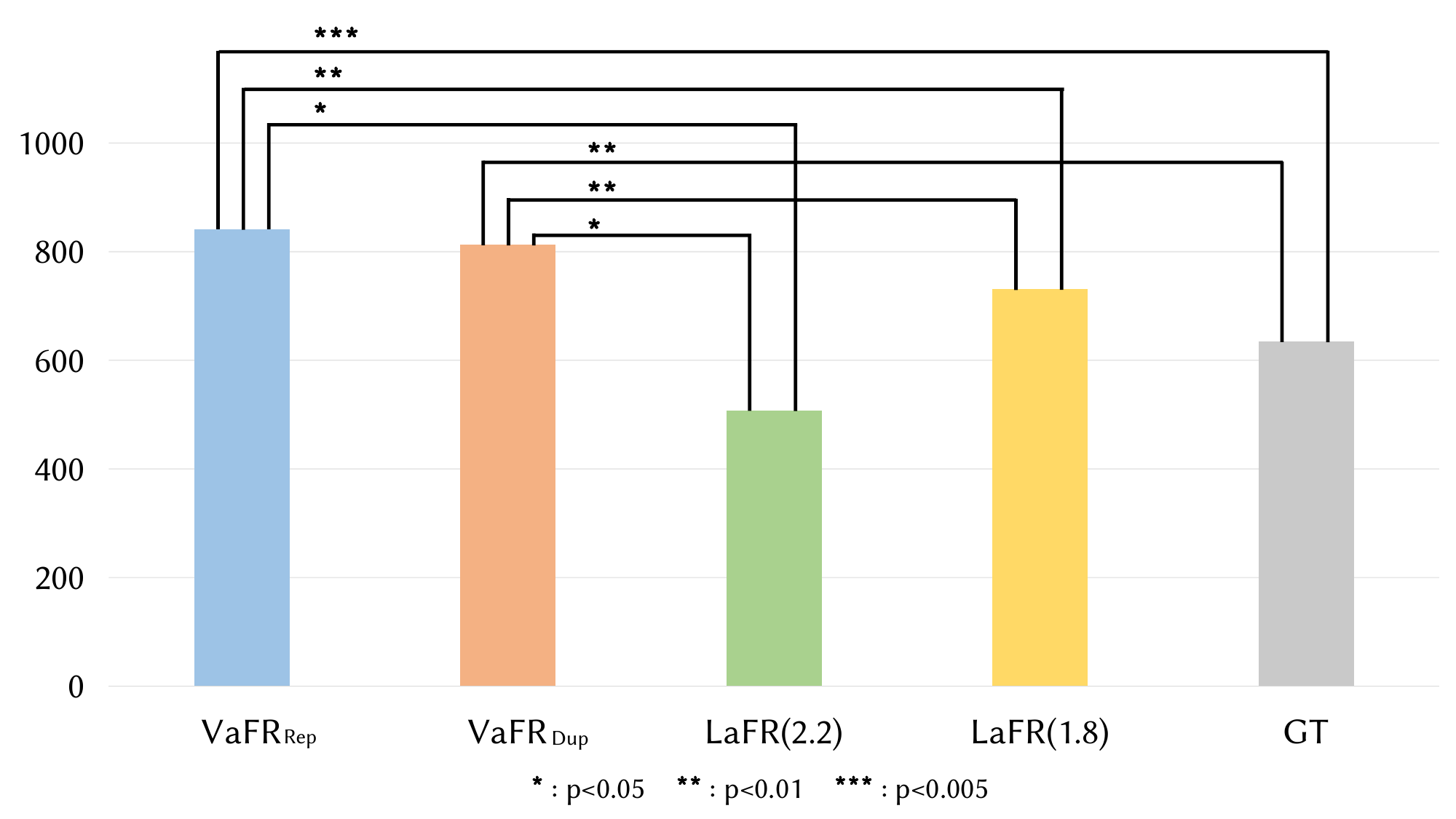}
	\caption{User preference of 2AFC experiment. VaFR$_\text{{Rep}}$ and VaFR$_\text{{Dup}}$ receive the most votes and consistently outperformed other solutions with statistical significance.}
	\label{fig:Preference}
\end{figure}

\begin{table}[t]  
\caption{Kruskal-Wallis test for pairwise comparisons within 2FAC experiment. Both VaFR$_\text{{Rep}}$ and VaFR$_\text{{Dup}}$ show p values smaller than 0.05 when compared to the alternatives, signifying that our approaches are significantly superior to these alternatives.}

\begin{center}     
\setlength\tabcolsep{2pt}
\begin{tabular}{c|c|c|c|c|c}   
\hline    P values                     & VaFR$_\text{{Rep}}$ & VaFR$_\text{{Dup}}$ & LaFR(2.2) & LaFR(1.8) & GT   \\ 
\hline   VaFR$_\text{{Rep}}$ &    /                        &0.522  &  0.004   & 0.016    & 0.006 \\  
\hline   VaFR$_\text{{Dup}}$                        &    /                        &  /   &  0.006   & 0.025    & 0.006 \\ 
\hline   LaFR(2.2)                     &    /                        &  /   &   /     & 0.873    & 0.631\\ 
\hline   LaFR(1.8)                     &    /                        &  /   &   /     &   /     & 0.749 \\ 
\hline   
\end{tabular}   
\end{center}
\label{table:Kruskal-Wallis} 
\end{table}

\subsubsection{Results and Analysis}
\label{userAnalysis}

We establish null hypotheses for the user study: There is no significant difference in user preference data for specific rendering modes between any two groups. We performed the Kruskal-Wallis test for pairwise comparisons at a significance level of 0.05, applying the Holm-Bonferroni correction. As indicated in \autoref{table:Kruskal-Wallis} and \autoref{fig:Preference}, the corrected p-values for VaFR$_\text{Rep}$ and VaFR$_\text{Dup}$ are below 0.05, rejecting the null hypothesis and indicating that our approach significantly outperforms alternative approaches. In contrast, the p values for LaFR(2.2), LaFR(1.8), and GT exceed 0.05, showing no significant differences among them. These findings are consistent with the results of the pilot study, confirming similar conclusions.
In all, VaFR$_\text{{Dup}}$ and VaFR$_\text{{Rep}}$ all offer superior perceptual outcomes compared to the other methods.

\begin{figure}[t]
	\centering 
	\includegraphics[width=0.9\columnwidth]{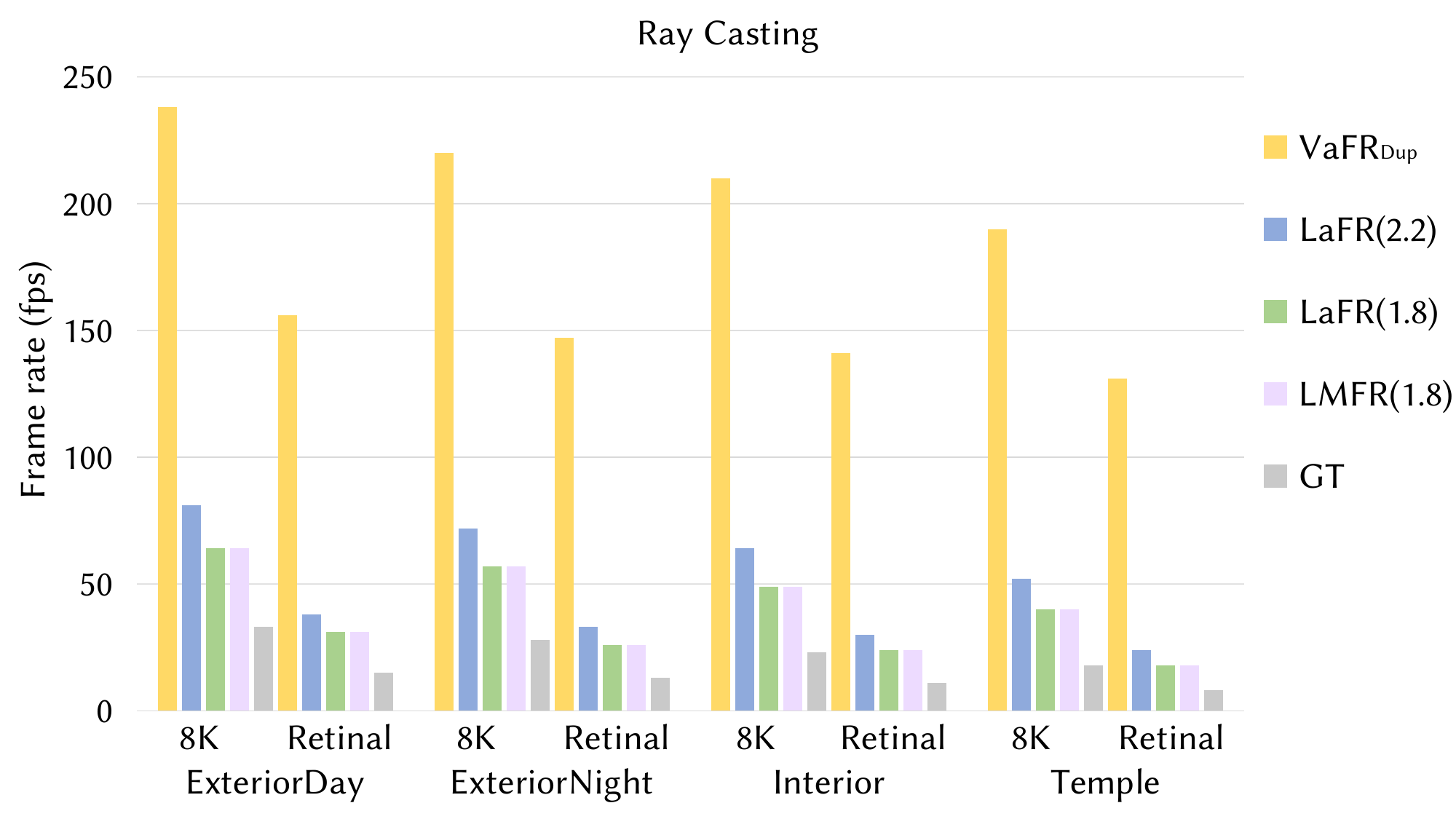}
	\caption{Performance comparison across different approaches are evaluated in various testing scenes, at both 8K and retinal resolutions. It reveals that VaFR$_\text{{Dup}}$ achieves superior performance, with over 131 fps at retinal resolution, while LaFR lags behind, remaining below 40 fps.}
	\label{fig:PerformanceRC}
\end{figure}

\begin{figure}[t]
	\centering 
	\includegraphics[width=0.85\columnwidth]{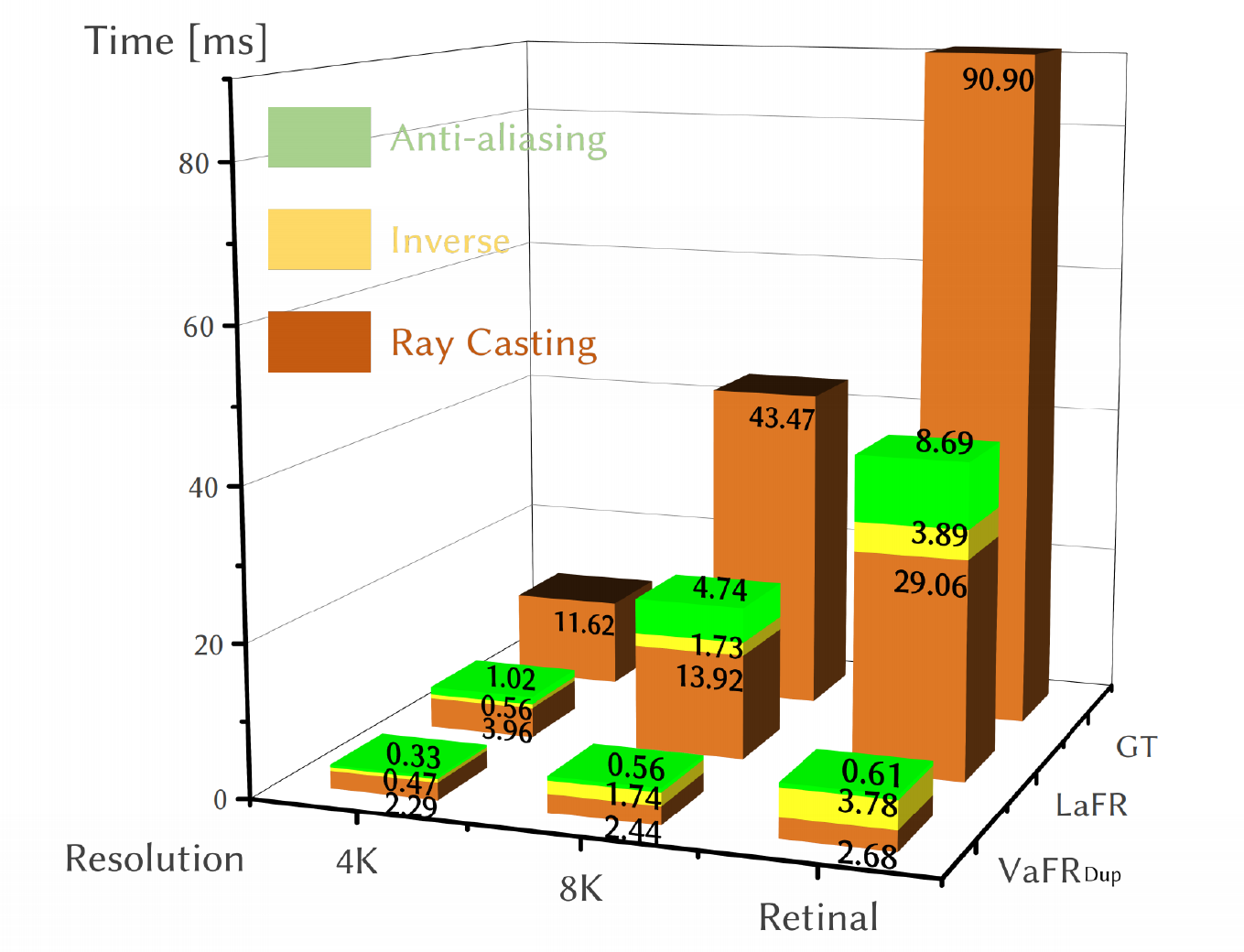}
	\caption{The average time cost of three stages for each approach across different resolutions.}
	\label{fig:times}
\end{figure} 

\begin{figure}[t]
	\centering 
	\includegraphics[width=0.95\columnwidth]{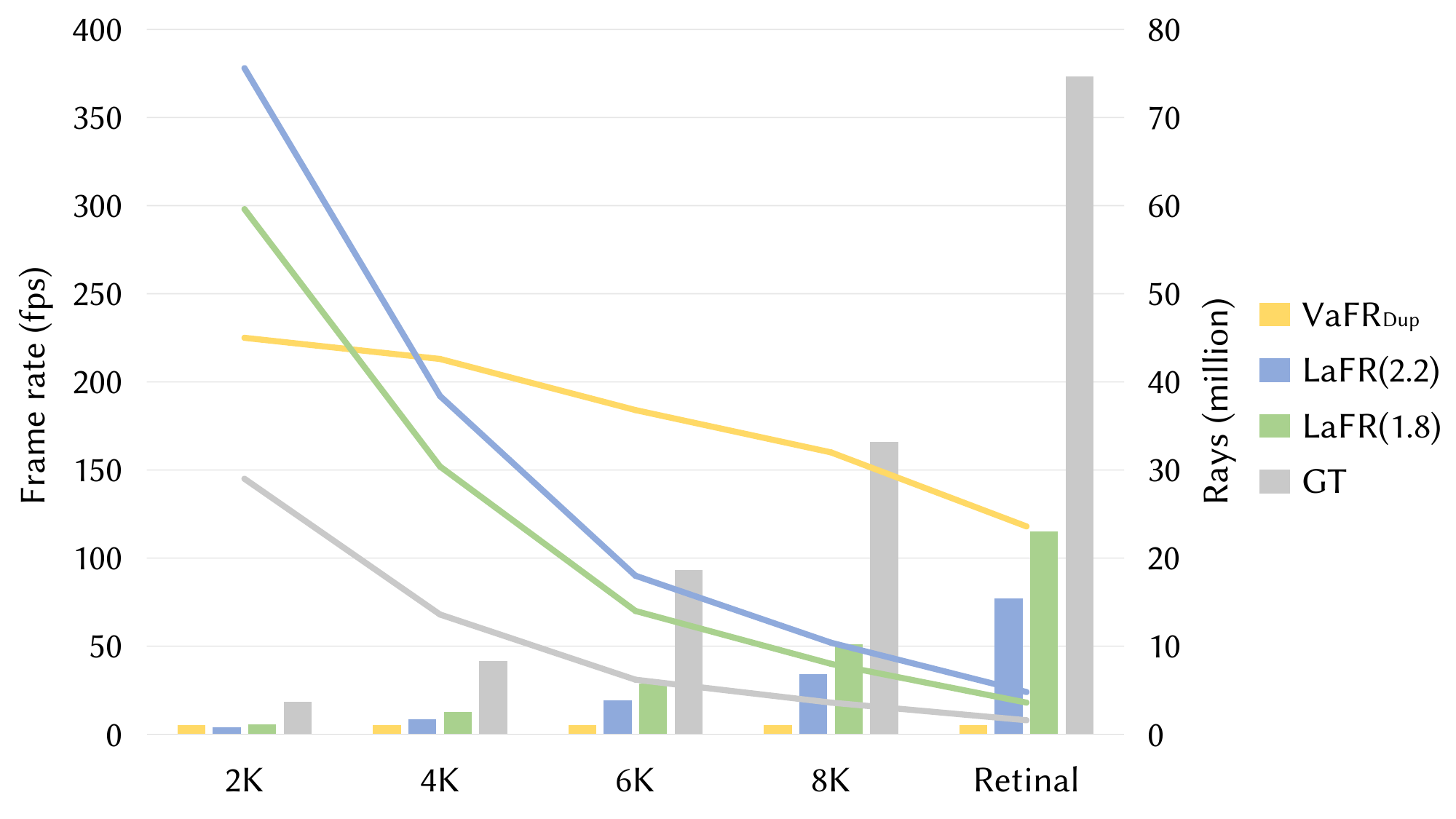}
	\caption{As resolution increases from 2K to retinal resolution, all methods experience a decline in performance. However, our approach maintains a more gradual decline, consistently achieving over 131 fps at retinal resolution, outperforming others. The histogram shows the number of rays emitted per eye.}
	\label{fig:DecreaseCurveRC}
\end{figure}

\begin{figure*}[t]
	\centering 
	\includegraphics[width=0.95\textwidth]{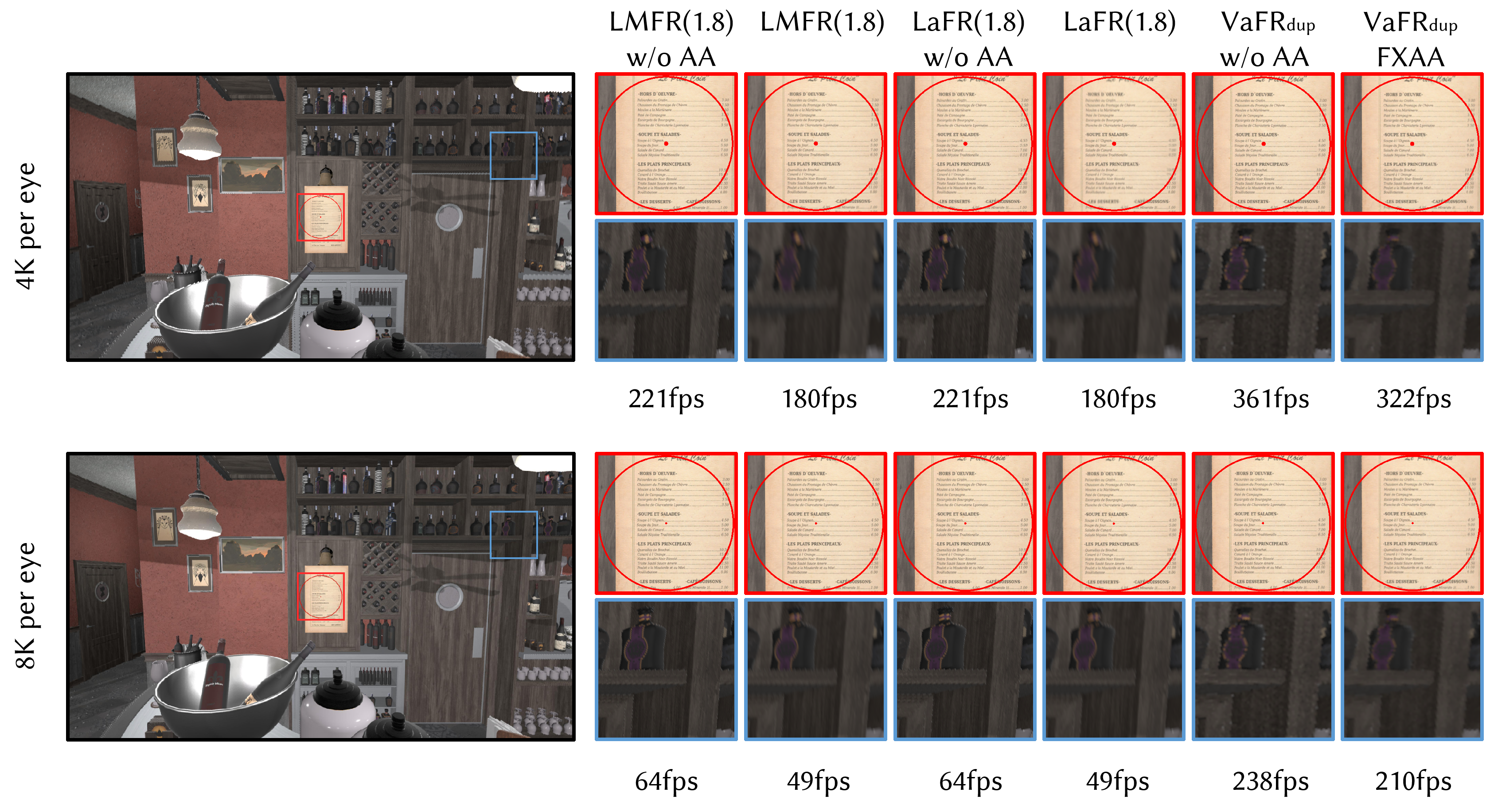}
	\caption{We compare image quality across three rendering algorithms, with and without anti-aliasing, at 4K and 8K resolutions, respectively. LMFR(1.8) and LaFR(1.8) use internal-AA and post-AA as described in \Cref{ImageQuality}. Zoomed-in results are cropped at eccentric angles of $0^\circ$ and $40^\circ$, respectively. Our method consistently delivers superior performance, with or without anti-aliasing, across various rendering resolutions, achieving higher performance without compromising perceptual visual quality.}
	\label{fig:FXAA}
\end{figure*}

\section{Validation on Ray Casting Pipeline}
\label{sec:ray casting}
To further demonstrate the compatibility and superiority of our algorithm, we evaluate the performance of our algorithm and the alternatives in the ray casting pipeline (without rasterization). 

\subsection{Ray Casting Pipeline}
In our new binocular rendering system for ray casting, a ray generation shader serves as the initial entry point to initiate ray casting. Texture and dispatch sizes are adjusted to match the constant LP buffer size. For each pixel in the LP buffer, its corresponding position in the screen space is calculated, and a ray is generated from this position. Shading occurs at the hit point using a closest hit shader, while a missing shader is applied when no geometry is hit. 
Upon successful triangle hit detection, the same shading code used in the fragment shader of the deferred rendering pipeline is executed. Following internal anti-aliasing, inverse mapping, and post anti-aliasing passes, the left image is generated. Subsequently, the right image is produced using the same set of passes as applied to the left image.

\subsection{Performance Evaluation}
Four scenes are tested using five rendering approaches, including VaFR$_\text{{Dup}}$, LaFR(2.2), LaFR(1.8), LMFR(1.8), and GT, all integrated into the ray tracing pipeline. The setup and procedure are the same as described in \Cref{setupandprocedure}.

\autoref{fig:PerformanceRC} illustrates average frame rates in four scenes for various rendering approaches at 8K per eye resolution and retinal resolution.
In the absence of rasterization bottlenecks in the ray casting pipeline, LMFR and LaFR achieve approximately a speedup of 2$\times$ in all scenes. Meanwhile, our method exhibits remarkable speedup, ranging from $7.2\times$ to $10.6\times$ across all scenes at 8K resolution, exceeding the range $1.9\times$ to $2.9\times$ observed in other approaches. Similarly, at retinal resolution, our method excels with speedup ranging from $10.4\times$ to $16.4\times$, maintaining frame rates between 131 and 156 fps, outperforming the range $2.0\times$ to $3.0\times$ seen in other methods. 

Compared to the deferred rendering pipeline that includes a rasterization stage, our method in the ray casting pipeline demonstrates an even more powerful ability to accelerate rendering. Our method's significant speedup advantage in the ray casting pipeline comes from a key process: emitting and shading rays from all pixels in the LP buffer, and then writing them back into it. This efficiency is further enhanced because our algorithm's LP buffer is sized based on the human visual acuity model, making it much smaller than typical display resolutions such as 8K. As a result, our approach maintains a smooth frame rate, demonstrating a powerful advantage in speedup.

\autoref{fig:times} shows the average time cost of the three stages for each approach at different resolutions within a single frame of the ray casting pipeline.
\autoref{fig:DecreaseCurveRC} depicts the frame rate fall-off as resolution increases from 2K to retinal level in the Temple scene. Despite the performance drop, VaFR$_\text{{Dup}}$ consistently maintains a frame rate of above 131 fps at retinal resolution, while others cannot.
A heavy rasterization workload can hinder rendering efficiency, whereas a ray-tracing pipeline fully exploits the advantages of our algorithm. In the ray casting pipeline, VaFR$_\text{{Dup}}$ achieves significantly higher frame rates at retinal resolution than other methods, effectively enabling retinal rendering in VR while ensuring a smooth visual experience.

\section{Ablation Study on Anti-Aliasing}
\label{ImageQuality}
Analysis of shading rate and performance plots reveals minimal differences among the comparative algorithms at a display resolution of around 4K. Therefore, we consider the significant impact of anti-aliasing algorithms on final image quality and frame rate through an ablation study on anti-aliasing. We compare image quality across various rendering modes within the ray casting pipeline at both 4K and 8K per eye resolution, as shown in \autoref{fig:FXAA}.

LMFR employs two anti-aliasing methods: internal anti-aliasing with a 3 $\times$ 3 Gaussian filter on the LP buffer, and post anti-aliasing with foveal-aware Gaussian filters on the final image. To ensure a fair comparison, we evaluate LMFR both with and without anti-aliasing, applying the same configurations to LaFR. Additionally, we compare the versions of VaFR$_\text{Dup}$ with and without FXAA.

Zoomed-in results are cropped at eccentric angles of $0^\circ$ and $40^\circ$ for a clearer quality comparison. The anti-aliasing methods used in LMFR offer only marginal improvement, as the Gaussian filter tends to blur the image. In contrast, applying FXAA to the LP buffer in our approach significantly enhances image quality with minimal impact on performance. This also demonstrates that our method consistently achieves superior performance, with or without anti-aliasing, across different rendering resolutions.
The results also show that LaFR exhibits slightly better quality outside the foveal region compared to VaFR$_\text{{Dup}}$; however, after applying internal anti-aliasing and post anti-aliasing, it falls slightly behind VaFR$_\text{{Dup}}$. Notably, at 8K per eye resolution, LaFR demonstrates significantly superior image quality outside the foveal region compared to VaFR$_\text{{Dup}}$, but its frame rate is only about one-quarter of VaFR$_\text{{Dup}}$'s.
This indicates that our method, VaFR, allocates computing resources more efficiently than the alternative, achieving higher performance without compromising perceptual visual quality.

\begin{figure*}[tb]
    \flushleft
	\includegraphics[width=0.95\textwidth]{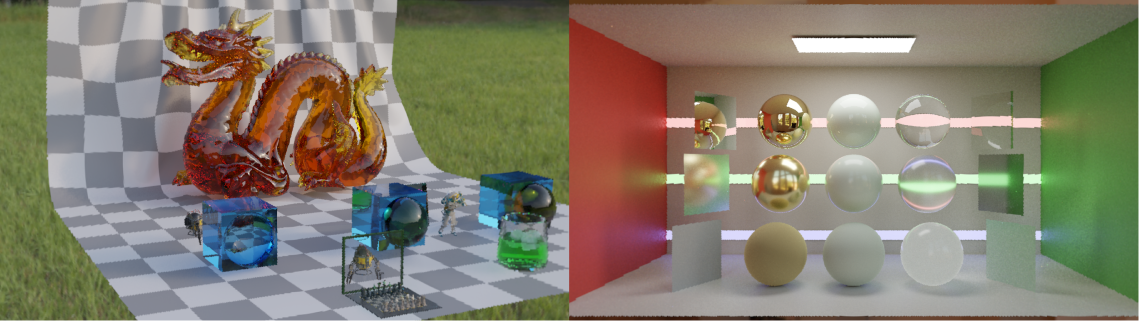}
	\caption{Path tracing examples (``Programmer-art" and ``Convergence-test") employing our approach, which renders at 8K per-eye resolution with a frame rate of 72 fps. The ROI is at the center of the image.}
	\label{fig:PT}
\end{figure*}

\section{Validation on Path Tracing Pipeline}

To validate the availability of our method to highly realistic rendering pipelines, we also integrated our approach into NVIDIA's PathTracing SDK \cite{Filip2023PTSDK} on a workstation featuring a robust 3.5 GHz Intel(R) Core(TM) i5-13600KF CPU, 32GB memory, and an NVIDIA GeForce GTX 4070Ti GPU.

\subsection{Path Tracing Pipeline}
In the ray tracing stage, mirroring our approach in the ray casting pipeline, we align the dispatch size and texture size with the dimensions of the LP buffer. Rays are then emitted from the screen-space position of each pixel in the LP buffer, producing a semi-elliptical-shaped image, as depicted in \autoref{fig:VaFR_framework}. Finally, in the inverse stage, the screen-space image is reconstructed.

Two typical scenes are tested, \emph{Programmer-art} and \emph{Convergence-test} from Path Tracing SDK \cite{Filip2023PTSDK}, configuring all settings to their default values, including 1 sample per pixel (spp), real-time noise, denoiser, 15 maximum bounces, 3 maximum diffuse bounces, and the use of Next Event Estimate. For simplicity, Russian Roulette, ReSTIR DI, and ReSTIR GI are disabled. We integrate our VaFR and LaFR(2.2) algorithms into the path tracing pipeline and make the necessary code modifications to output images at 8K resolution.

\subsection{Performance Evaluation}
As shown in \autoref{fig:PT}, the gaze point is positioned in the center of the screen. It can be observed that the images near the gaze point are noticeably clear, while the shading rate falls in the peripheral field of view. From this we can conclude that our algorithm remains applicable to path tracing, a rendering method known for its high realism, showcasing strong universality.
 
At a monocular 8K resolution, VaFR$_\text{{Dup}}$ achieves a frame rate of approximately 144 fps, while LaFR(2.2) reaches only 14 fps. When rendering for VR with binocular setting, we reasonably estimate that VaFR$_\text{{Dup}}$ has a smooth frame rate of around 72 fps, whereas LaFR(2.2) lags at 7 fps. In comparison, ground truth (GT) real-time rendering at 2K can only achieve a binocular frame rate of 15 fps, and rendering at 8K will lead to system crashes. 

Achieving 72 fps in 8K binocular rendering already approximates the initial perceptual requirements of VR users, though it falls slightly short of the commonly accepted comfort frame rate of 90 fps. However, this gap is not unbridgeable; a more advanced GPU can further improve the frame rate. This performance indicates that our VaFR significantly enhances path tracing in a high-resolution VR headset, demonstrating considerable potential for highly realistic rendering.

\section{Further Discussion of Anisotropic Visual Acuity}
\label{Anisotropic}

\begin{figure}[tb]
	\centering 
	\includegraphics[width=0.95\columnwidth]{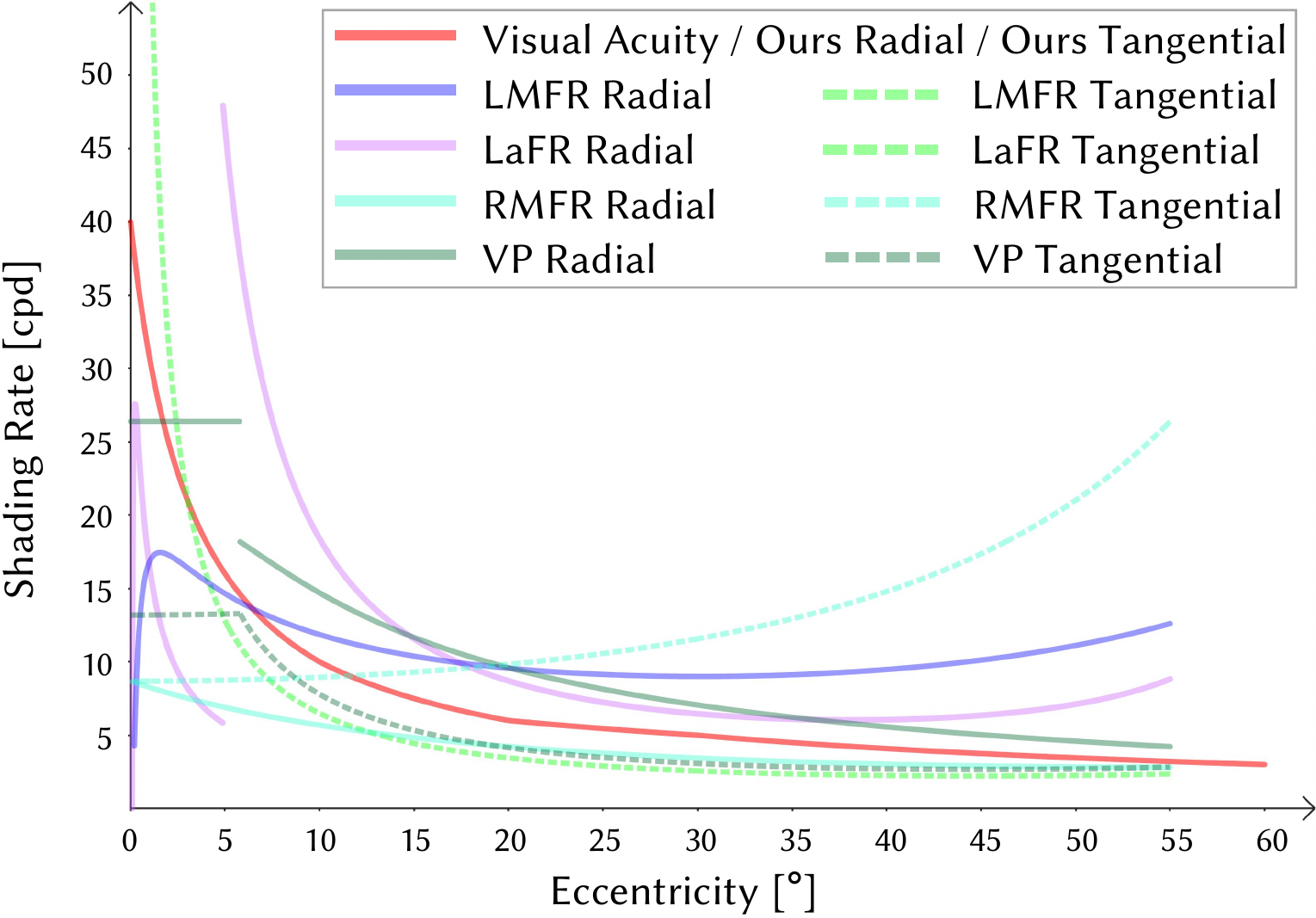}
	\caption{Our algorithm precisely aligns radial and tangential shading rates with human visual acuity, ensuring consistent performance under all conditions. In contrast, other methods often exhibit deviations in similar scenarios.}
	\label{fig:OurRadialTangential}
\end{figure}

Humans have anisotropic visual acuity, with detection and resolution varying between frequencies in radial and tangential directions. We are more sensitive to radially oriented gratings \cite{Rovamo:1982}, approximately twice as much at $30^\circ$ eccentricity \cite{Thibos:1996}. Although our primary focus does not revolve around this particular issue, it is worth noting that our approach has the inherent advantage of independently adjusting tangential shading rates.

In \Cref{section:Formulation}, we keep the tangential shading rate equal to the radial shading rate as \autoref{func:tangential} for simplicity. If we set the ratio factor of tangential to radial as $\Delta(e)$ that varies with the eccentricity angle $e$, \autoref{func:v(e,theta)} only needs to multiply by it as: 
\begin{equation}
	v(e,\theta) = \Delta(e)\cdot\theta \cdot \frac{\sin{2e}}{me+\omega_0}\  .
\label{func:v(e,theta,Delta)}
\end{equation}
This conveniently enables the incorporation of anisotropy for additional performance gains, which is an aspect not addressed by previous foveated rendering algorithms.

As illustrated in \autoref{fig:OurRadialTangential}, both LMFR(1.8) and LaFR(1.8) use the same parameter $\delta=1.8$, resulting in tangential shading rate curves that perfectly coincide. 
Their tangential shading rates are calculated as: 
\begin{equation}
	SR_{tangential}(e) = \frac{{h}\slash{1.8}}{2\pi r\frac{\mathrm{d}e}{\mathrm{d}r}} = \frac{{h}\slash{1.8}}{720\cos{e}\sin{e}}\ ,
\label{func:TheirSRTan}
\end{equation}
where $h$ represents display resolution height. As depicted in green, these tangential shading rates deviate from both the visual acuity and their radial shading rates. This discrepancy is unreasonable.

Referring to \autoref{func:tangential}, we infer that our radial and tangential shading rates are equivalent, resulting in identical curves.
Our algorithm exhibits a significantly reasonable tangential shading rate and can be independently tuned.

\section{Conclusions, Limitations, and Future Work }
\label{limitation}

We have presented a visual acuity consistent mapping approach for highly efficient foveated rendering, which is applicable to various binocular rendering strategies in VR settings. Our approach exhibits promising performance at high resolutions while maintaining a better perceptual quality than competing methods. With a deferred rendering pipeline, our approach achieves up to $7.6\times$ speedup over GT method, reaching 61 fps at 8K per-eye resolution. Using a ray casting pipeline, our approach also delivers an impressive   10.4$\times$ to 16.4$\times$ speedup compared to the GT method at retinal resolution, achieving frame rates between 131 and 156 fps. Additionally, our method enhances path tracing performance, boosting the frame rate of binocular 8K rendering to approximately 72 fps, enabling highly realistic VR rendering.

Our approach still has some limitations. Temporal flickering, as a known issue associated with the log-polar mapping method, still requires further investigation. As stated in Meng et al. \cite{Meng:2018:Kernel}, ``However, in fly-through of the scene with glossy objects, we notice that view-dependent specular reflection changes before and after applying KFR. Foveation amplifies the specular reflection regions, and makes the specular highlight flicker more." Foveation exacerbates these specular reflection variations, causing specular highlights to flicker more. Without reprojection as the second stage, the log-polar mapping method essentially re-renders the right-eye image, and due to disparities in camera position and gaze orientation, log-polar mapping may introduce slight mismatches in specular highlights between the left and right views. Using reprojection as the second stage helps mitigate this issue by addressing the mismatch between the left and right views.

Additionally, Patney et al. \cite{Patney:2016} developed a saccade-aware multi-resolution temporal anti-aliasing method, enhancing peripheral details through contrast enhancement. Tariq et al. \cite{Tariq:2022} introduced a perceptual technique, substituting indiscernible frequencies with procedural noise, customized to image and perception specifics. These post-processing methods, compatible with our approach that provides screen-space images, can be seamlessly integrated into our algorithm to heighten efficiency.

Another challenge associated with foveated rendering is the Moire patterns, particularly in approximately $30^{\circ}$ peripheral regions with a lower spatial sampling frequency. The spatial sampling frequency that aligns with the spatial stimuli frequency of the scene will lead to the appearance of Moire patterns. Notably, these patterns evolve with changes in gaze point or head movement. While the foveal and far peripheral regions are less susceptible to Moire patterns, addressing this issue is still essential.

Furthermore, implementing physical brightness-based edge detection and smoothing represents an important challenge for our future research. Anti-aliasing in the LP buffer significantly enhances the final image quality. To achieve this, we use FXAA to smooth the edges. However, tone-mapping within the log-polar space before FXAA poses challenges, as average luminance calculations are not straightforward in this space. Consequently, alternative physical-based edge detection and smoothing methods are necessary.


\bibliographystyle{abbrv-doi}
\bibliography{my}

\end{document}